\documentclass[twocolumn,hyperpdf,amsmath,amssymb,aps,prd,10pt,superscriptaddress,nofootinbib,noeprint,preprintnumbers,floatfix]{revtex4-1}

\usepackage{amsmath} 
\usepackage{amssymb} 
\usepackage{physics} 
\usepackage{graphicx}
\usepackage{xcolor}
\usepackage[caption=false]{subfig}

\usepackage{hyperref} 

\usepackage{mathtools, multirow}

\usepackage[letterspace=-10]{microtype}

\begin{document}

\title{Timelike meson form-factors beyond the elastic region from lattice QCD} 

\preprint{JLAB-THY-24-4132}

\author{Felipe G.\ Ortega-Gama}
\email[]{felortga@jlab.org}
\affiliation{Department of Physics, William \& Mary, Williamsburg, VA 23187, USA}

\author{Jozef J.\ Dudek}
\email[]{dudek@jlab.org}
\affiliation{Department of Physics, William \& Mary, Williamsburg, VA 23187, USA}
\affiliation{\lsstyle Thomas Jefferson National Accelerator Facility, 12000 Jefferson Avenue, Newport News, VA 23606, USA}

\author{Robert G.\ Edwards}
\affiliation{\lsstyle Thomas Jefferson National Accelerator Facility, 12000 Jefferson Avenue, Newport News, VA 23606, USA}

\collaboration{for the Hadron Spectrum Collaboration}

\date{\today}

\begin{abstract}
We present a calculation of the vector-isovector timelike form-factors of the pion and the kaon using lattice quantum chromodynamics. We calculate two-point correlation functions with \mbox{$m_\pi \sim 280$ MeV}, extracting both the finite-volume spectrum and matrix elements for these states created from the vacuum by a vector current. After determining the coupled-channel $\pi\pi, K\overline{K}$ scattering amplitudes, we perform the necessary correction for the significant finite-volume effects present in the current matrix elements, leading to the timelike form-factors. We find these to be dominated by the presence of the $\rho$ resonance, and we extract its decay constant by an analytic continuation of the amplitudes to the resonance pole. In addition, the spacelike pion form-factor is determined on the same lattice configurations, and a dispersive parameterization is used to simultaneously describe the spacelike and elastic timelike regions.
\end{abstract}

\maketitle

\section{\label{sec:intro}Introduction}

The lightest octet of pseudoscalar mesons plays a unique role in QCD, being bound states of quarks and gluons, while also serving as the pseudo-Goldstone bosons of broken chiral symmetry. How this is manifested can be explored via a comprehensive description of their internal structure, and this can advance our understanding of the non-perturbative regime of QCD.
Quantitative measures of internal structure can come from transitions mediated by external probes such as electroweak currents; and one of the simplest transitions, which is widely studied both theoretically and experimentally, is the vector form-factor of the pion, $f_\pi(P^2)$, corresponding to the interaction vertex between a photon of virtuality $P^2$ and two charged pions. Depending upon the external kinematics, this can describe two kinematically disjoint processes: for \emph{spacelike} $P^2<0$ it corresponds to the elastic electromagnetic response of a $\pi^\pm$ meson, while for \emph{timelike} $P^2>4m_\pi^2$ it describes the production amplitude of a $\pi^+\pi^-$ pair from the QCD vacuum.
Experimentally, the spacelike form-factor is extracted from elastic $e^- \pi^\pm$ scattering, either from a pion beam on atomic electrons~\cite{NA7:1986vav}, or electroproduction of charged pions from a nucleus~\cite{JeffersonLab:2008jve}, while the timelike form-factor can be retrieved from the $e^+e^-\to \pi^+ \pi^-$ cross-section~\cite{Fang:2021wes}, or the tau-lepton decay rate, $\tau^-\to \pi^-\pi^0\nu_\tau$~\cite{Belle:2008xpe}. Ref.~\cite{Colangelo:2018mtw} presents a recent comprehensive review. Below the $K\overline{K}$ threshold, the largest contribution to the timelike pion form-factor comes from the $\rho$ resonance which generates a broad peak in $f_\pi(P^2)$. Once past the $K\overline{K}$ threshold, $P^2 > 4 m_K^2$, the production process becomes inelastic, and the amplitude describing $\gamma \to K\overline{K}$ can be expressed in terms of the kaon timelike form-factor.

Theoretical calculation of the pion form-factor from first principles in QCD is possible within lattice QCD, a systematically improvable numerical implementation of the QCD path integral considered on a discrete grid of space-time points of finite extent. Owing to the Euclidean signature used in this approach, real-time dynamics like scattering are not directly accessible, but the discrete spectrum of finite-volume states and local matrix elements can be extracted from correlation functions.
The form-factor in the spacelike region can be obtained from matrix elements directly accessible from three-point correlation functions having only exponentially suppressed finite-volume corrections, with many previous calculations being reported (a summary with references can be found in Ref.~\cite{FlavourLatticeAveragingGroupFLAG:2021npn}).  
In the timelike region, on the other hand, the vector current produces $\pi\pi$ pairs which in infinite volume strongly rescatter, and in finite volume these pions can go on-shell and sample the boundary of the lattice, leading to significant multiplicative finite-volume corrections to the matrix elements extracted from two-point correlation functions. Evaluation of these correction factors requires knowledge of the meson-meson scattering amplitude with vector quantum numbers, but this can be determined applying the L\"uscher approach~\cite{Luscher:1985dn,Luscher:1986pf} to the discrete spectrum of states extracted from a matrix of two-point correlation functions, a technique that is now well established even in the case of coupled-channels~\cite{Briceno:2017max}.

\medskip

The finite-volume correction factor required in production amplitudes follows from the pioneering work of Lellouch and Lüscher~\cite{Lellouch:2000pv} with the case of the timelike form-factor of the pion presented in Ref.~\cite{Meyer:2011um}. The formalism has been applied in explicit lattice QCD calculations, restricted to the elastic region of $\pi\pi$ production, in Refs.~\cite{Feng:2014gba,Andersen:2018mau,Erben:2019nmx}.
Many more lattice calculations of $\pi\pi$ scattering have been performed, which is a prerequisite for the form-factor volume corrections, see Ref.~\cite{Mai:2022eur} for a recent review of these.

In this paper, we will perform a lattice QCD calculation extended into the inelastic energy region where production of $K\overline{K}$ as well as $\pi\pi$ is possible, showing that the appropriate finite-volume corrections can be applied in this coupled-channel situation. In addition, we will compute the spacelike pion form-factor on the same lattice configurations, and present amplitudes with appropriate analytic properties which describe the form-factor across both spacelike and elastic timelike regions.

\section{Meson-meson production} \label{sec:production}

\begin{figure*}
\subfloat[\label{fig:diag_sl_pi}Spacelike $\pi$ form factor.]{\includegraphics[width=.2\linewidth]{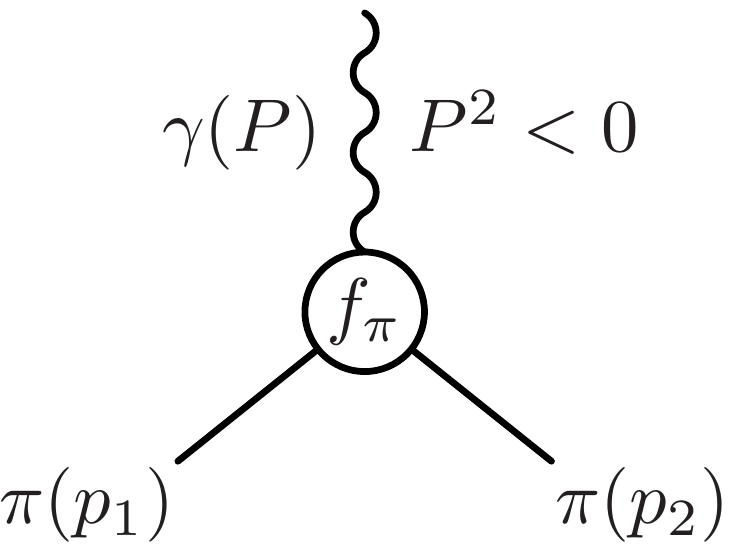}}
\hfil
\subfloat[\label{fig:diag_tl_pi}Timelike $\pi$ form factor.]{\includegraphics[width=.23\linewidth]{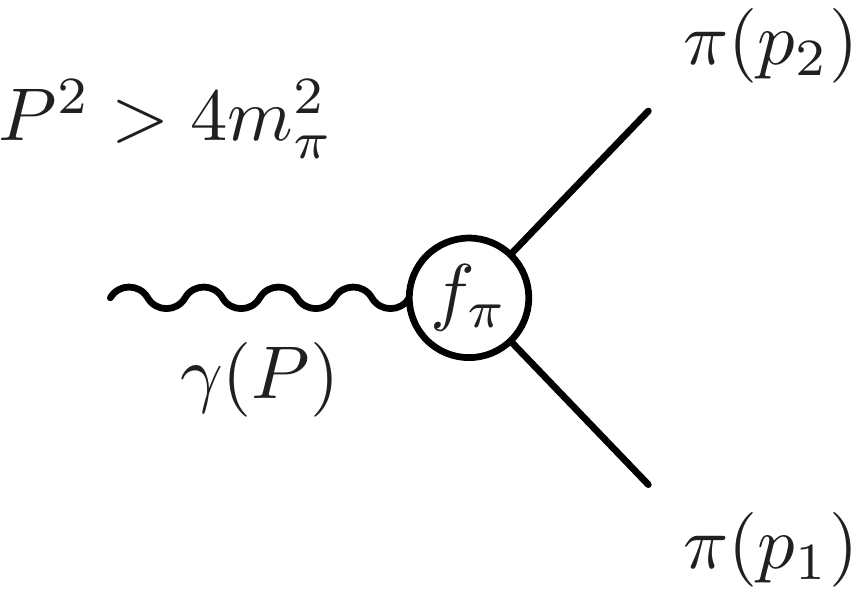}}
\hfil
\subfloat[\label{fig:diag_tl_K}Timelike $K$ form factor.]{\includegraphics[width=.23\linewidth]{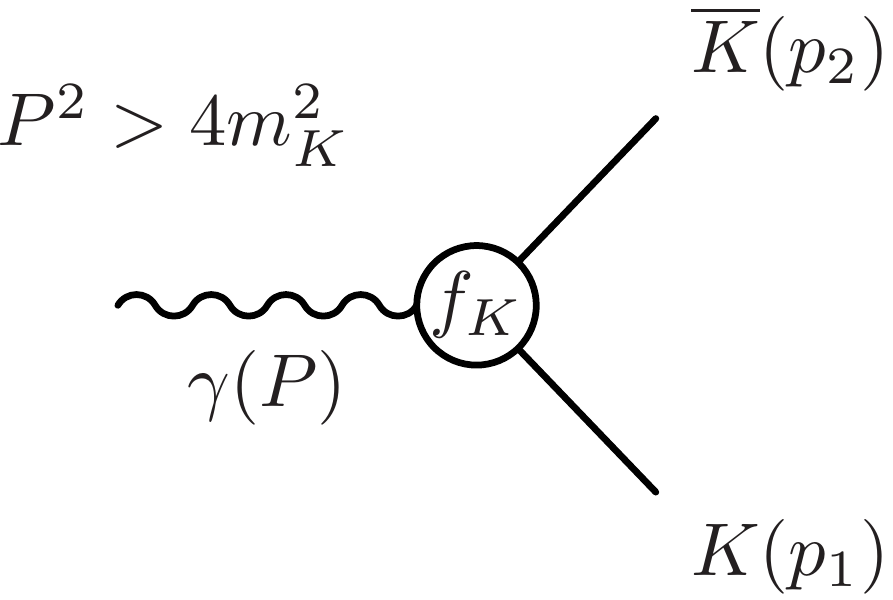}}
\caption{\label{fig:diags}Diagrams representing the processes considered in this paper.
The spacelike and timelike form factors of the pion are aspects of a single function of the photon virtuality, $f_\pi(P^2)$. }
\end{figure*}

The amplitude describing the production of a pair of pseudoscalars from the vacuum by a conserved vector current, $\mathcal{J}^\mu$, can be decomposed into a kinematic factor multiplying a single Lorentz-invariant form factor,
\begin{align}
\mathcal{H}_a^\mu (p_1,p_2)&\equiv \mel{p_{1},p_{2}; a}{\mathcal{J}^\mu(x=0)}{0} \nonumber \\
&= (p_1-p_2)^\mu \, f_{a}(s)\,,
\end{align}
where $a$ indicates the pair ($\{\pi\pi, K\overline{K}, \ldots\}$) and $p_{i=1,2}$ the four-momentum of each pseudoscalar meson.
The form factors, $f_a(s)$, are expressed in terms of the square of the center-of-momentum (COM) frame energy, ${s=(p_1+p_2)^2}$, which corresponds to the \emph{timelike} virtuality, $s>0$, of the current in the $s$-channel production.
The crossed-channel process in which a pion (or kaon) absorbs an off-shell current is described by the same form factor for \emph{spacelike} virtualities, $s<0$, as indicated in Fig.~\ref{fig:diags}.

While this amplitude decomposition by construction describes only a single $J^P=1^-$ partial-wave, explicit partial-wave projection into definite spin-projection, $m$, brings it into a form more natural for application of the finite-volume formalism,
\begin{equation*}
    \mathcal{H}^\mu_{a, m}(P) = \int\! \frac{\dd \Omega^\star}{\sqrt{4\pi}}\,  Y^*_{1 m}( \Omega^\star ) \, 
\mathcal{H}^\mu_a(p_1,p_2)\,,
\end{equation*}
in a frame with four-momentum $P^\mu= p_1^\mu+p_2^\mu$. Evaluating the integral yields
\begin{equation}\label{eq:pw_kin}
    \mathcal{H}_{a,m}^\mu(P) = K^\mu_m \, k_a^\star(s) \, f_a(s)\,  ,
\end{equation}
with $K^\mu_m = \sqrt{\tfrac{4}{3}}\, \epsilon^{\mu*}(P, m)$, which features the polarization vector, and, characteristic of a $P$-wave process, one power of $k^\star_a(s)$, the magnitude of the momentum of a pseudoscalar in the COM-frame.

As is also the case for the scattering $S$-matrix describing the hadronic processes $\pi\pi \to \pi\pi, \pi \pi \to K\overline{K}, \ldots$, the form factors are constrained by the principles of unitarity, causality, and crossing-symmetry.
Of these, unitarity has the simplest implementation, and the most direct consequence -- it can be expressed as a constraint on the imaginary part of the amplitude leading to
\begin{equation}\label{eq:Imf}
k_a^\star(s)\, \text{Im} f_a(s) = \sum\nolimits_b\,  \mathcal{M}_{ab}^*(s)\, \rho_b(s)\, k_b^\star(s)\,  f_b(s)\,,
\end{equation}
where the sum runs over all kinematically open pair channels at COM-frame energy $\sqrt{s}$. The presence of the \mbox{$P$-wave} hadronic scattering matrix, $\mathcal{M}$, in this expression can be interpreted as reflecting the strong hadron rescattering that must occur after a meson-meson pair is produced by the current.
The corresponding unitarity constraint on $\mathcal{M}$ reads,
\begin{equation}\label{eq:ImM}
\text{Im} \, \mathcal{M}_{ac} = \sum\nolimits_b \mathcal{M}_{ab}^*\, \rho_b\, \mathcal{M}_{bc}\,,
\end{equation}
where the presence of the phase-space for each channel, $\rho_a(s) = \tfrac{k_a^\star(s)}{8\pi\sqrt{s}}$, indicates the unitarity-enforced analytic structure with a branch cut opening at each new threshold.
Moving down from real energy values through any one of these cuts leads to a neighbouring Riemann sheet, on which pole singularities may appear, reflecting the presence of \emph{resonances}. For the case of $\pi\pi$ scattering, this is how the $\rho$ appears mathematically.

The similarity in structure between Eqs.~\ref{eq:Imf} and \ref{eq:ImM} indicates that the imaginary parts of the functions $f_a(s)$ and $\mathcal{M}_{ab}(s)$ are related, and in the case of \emph{elastic} scattering, this is the origin of Watson's theorem, which states that the \emph{phase} of $f(s)$ must be equal to that of $\mathcal{M}$, i.e.\ the elastic scattering phase-shift.
In the more general coupled-channel case, a parameterization which trivially satisfies Eq.~\ref{eq:Imf} is,
\begin{equation}\label{eq:fNovD}
f_a = \sum\nolimits_b \tfrac{1}{k_a^\star} \mathcal{M}_{ab} \tfrac{1}{k_b^\star}\,   \mathcal{F}_b \,,
\end{equation}
where functions, $\mathcal{F}_a(s)$, have been introduced which do not have the branch cut structure present in $\mathcal{M}(s)$, but rather are smooth, real functions in the kinematically accessible scattering region.
The presence of explicit factors of $k_a^\star(s)$ reflects the $P$-wave nature of both production and scattering processes.

Analyticity of the production and scattering amplitudes, which is related to both unitarity and causality, is less straightforward to implement as a constraint on explicit parameterizations.
It can be introduced by means of Cauchy's integral formula applied paying attention to the allowed singularities of the amplitude, such as the branch-cut discontinuities imposed by unitarity, and any bound-state poles.
The resulting integrals, known as dispersion relations, typically relate the real part of amplitudes at any energy to integrals over semi-infinite energy regions of their imaginary part, modulo some known kernel function.

In the case that scattering is treated as being elastic at all energies, a solution of the dispersion relations known as the Omnès--form exists~\cite{Omnes:1958hv}, which suggests the following decomposition for the elastic form factor,
\begin{equation}\label{eq:fOmnes_el}
f(s) = \Omega(s) \,\mathcal{F}_\Omega(s)\,,
\end{equation}
where the Omnès factor, $\Omega(s)$, replaces the factor $\mathcal{M}/k^{\star2}$ of the elastic version of Eq.~\ref{eq:fNovD}.
The Omnès factor has the same phase as the scattering amplitude, $\mathcal{M}$, as dictated by unitarity, but its magnitude is determined by a dispersion relation -- an explicit form will be presented later in this manuscript. The function $\mathcal{F}_\Omega(s)$ in Eq.~\ref{eq:fOmnes_el} is a smooth real function particular to the pion form-factor. 

The analytic properties of the Omnès factor, which by construction features no singularities for $s<0$, makes Eq.~\ref{eq:fOmnes_el} applicable in \emph{both} the timelike and spacelike regions, while to use Eq.~\ref{eq:fNovD} in the spacelike region would require the extrapolation of $\mathcal{M}(s)$ into a region where crossed-channel singularities appear,\footnote{Although in practical parameterizations of $\mathcal{M}(s)$ we may choose not include them when considering only energy regions away from their influence.} and these would need to be cancelled by singularities in $\mathcal{F}(s)$.

The Omnès form presented above is restricted to \emph{elastic} scattering, and the extension to the coupled-channel case is not so simple, with a set of coupled integral equations known as the Muskhelishvili--Omnès problem needing to be solved. For the case presented in this paper, there is limited benefit to such an undertaking, as we will see later.

\section{Finite volume formalism} \label{sec:fv}

Lattice QCD calculations necessarily work in a finite spatial volume, and as such do not feature a continuum of multiparticle states, but rather a discrete spectrum sensitive to the volume. 
For energies below three-particle thresholds, the finite-volume spectrum is related to two-particle scattering amplitudes by the  L\"uscher determinant condition~\cite{Luscher:1985dn,Luscher:1986pf,Luscher:1990ux,Luscher:1991cf,Rummukainen:1995vs,He:2005ey,Kim:2005gf,Christ:2005gi,Fu:2011xz,Guo:2012hv,Hansen:2012tf,Briceno:2012yi,Gockeler:2012yj,Leskovec:2012gb,Briceno:2014oea, Briceno:2017max},
\begin{equation}\label{eq:Luscherqc}
	\det \big[ \mathcal{M}(E) + F^{-1}(E,L) \big]= 0\,,
\end{equation}
where $F$ is a matrix of known functions of essentially kinematic origin dependent on the $L\times L \times L$ volume of the periodic lattice, and $\mathcal{M}$ is a matrix containing scattering amplitudes, diagonal in partial-waves, but dense in channel-space when coupled-channels are kinematically accessible.
The finite-volume spectrum, $\{E_\mathfrak{n}(L)\}$, corresponds to the discrete set of solutions to this equation for a given $\mathcal{M}(E)$, and as such, in general, finite-volume eigenstates, $\ket{\mathfrak{n}}_L$, cannot be associated with a particular channel or partial-wave.

We respect the cubic symmetry of the spatial lattice boundary by computing spectra in irreducible representations, `\emph{irreps}', of the reduced rotational symmetry group, and these contain subductions of multiple values of angular momentum, leading to the partial-wave space in the determinant condition. More constraint on scattering amplitudes can be obtained by computing spectra in frames in which the two-particle system moves relative to the fixed lattice, and in this case irreps of the \emph{little group} that preserves rotations of the cube around the direction of motion are used.

Parameterizing the energy dependence of partial-wave scattering amplitudes, $\mathcal{M}(E)$, we can solve Eq.~\ref{eq:Luscherqc} for volumes and irreps in which we have computed the lattice QCD spectrum, to find `model' spectra. Free parameters in the amplitudes can then be adjusted to bring the model spectra into agreement with the computed spectra, thus providing hadron scattering amplitudes constrained by QCD dynamics. This approach has been successfully applied to several systems of coupled-channel scattering, see Refs.~\cite{Dudek:2014qha, Wilson:2014cna, Wilson:2015dqa, Dudek:2016cru, Moir:2016srx, Briceno:2017qmb, Woss:2019hse, Woss:2020ayi,BaryonScatteringBaSc:2023ori, Prelovsek:2020eiw, Wilson:2023hzu}, with an efficient approach to solving Eq.~\ref{eq:Luscherqc} in the coupled-channel case presented in Ref.~\cite{Woss:2020cmp}. The approach is reviewed in Ref.~\cite{Briceno:2017max}.

\smallskip 
\emph{Production} amplitudes where a two-particle state is generated by the action of a current on the vacuum (or on a single-particle state), as introduced in the previous section, can be accessed by computing current matrix elements featuring in the initial or final state the discrete finite-volume states discussed above. The relationship of these finite-volume matrix elements to the infinite volume amplitudes at the same energy is given by,
\begin{multline}\label{eq:LLform}
	\big| \mel{\mathfrak{n}}{\mathcal{J}^\mu(x=0)}{0}_L \big|^2 =\\
     \tfrac{1}{2E_\mathfrak{n}L^3}\, 
\sum\nolimits_{a,b}
\mathcal{H}^\mu_a(E_\mathfrak{n}) \,\widetilde{\mathcal{R}}_{a,b}(E_\mathfrak{n}, L) \, \mathcal{H}^\mu_b(E_\mathfrak{n})\,,
\end{multline}
where the matrix $\widetilde{\mathcal{R}}$, the coupled-channel generalization of the Lellouch-L\"uscher factor~\cite{Lellouch:2000pv}, is related to the residue of the finite-volume two-particle propagator at the energy, $E_\mathfrak{n}(L)$, where the propagator has a pole~\cite{Briceno:2014uqa, Briceno:2015csa}.
The explicit form of the matrix $\widetilde{\mathcal{R}}$ is,
\begin{equation*}
	\widetilde{\mathcal{R}}(E_\mathfrak{n}, L) = 2E_\mathfrak{n}\cdot \lim_{E\to E_\mathfrak{n}} \frac{E-E_\mathfrak{n}}{\mathcal{M}(E)+F^{-1}(E,L)}\,,
\end{equation*}
where the matrix in the denominator can only have a single vanishing eigenvalue at $E_\mathfrak{n}$ in order to generate \emph{simple poles} in correlation functions, consistent with causality.
This means that $\widetilde{\mathcal R}$ is a rank-one matrix, which can be decomposed in terms of an outer product featuring the eigenvector associated with the vanishing eigenvalue, although in practice it is more convenient~\cite{Briceno:2021xlc} to use a matrix $\mathcal{M} \widetilde{\mathcal{R}} \mathcal{M}$, 
\begin{align} \label{eq:MRM}
	\mathcal{M}\,\widetilde{\mathcal{R}}\,\mathcal{M} &= -2E_\mathfrak{n} \cdot \lim_{E\to E_\mathfrak{n}} \frac{E-E_\mathfrak{n}}{\mathcal{M}^{-1}(E)+F(E,L)}\,, \nonumber \\
&= -\frac{2E^\star_\mathfrak{n}}{\mu^{\prime\star}_0} \, \mathbf{w}^{\phantom{\intercal}}_0\, \mathbf{w}_0^{\intercal} 
\,,
\end{align}
where $\mu_0^{\prime \star}$ is the slope in energy of the vanishing eigenvalue of $\mathcal{M}^{-1} + F$ at $E_\mathfrak{n}$, which is seen to control the normalization, while the eigenvector, $\mathbf{w}_0$, distributes strength across the channel (and/or partial-wave) space. 

Once $\widetilde{\mathcal{R}}$ is factorized in this way, Eq.~\ref{eq:LLform} becomes a linear relationship between the finite-volume and infinite volume amplitudes, and the introduction of explicit factors of $\mathcal{M}$ in Eq.~\ref{eq:MRM} removes the strong rescattering factor in Eq.~\ref{eq:fNovD} leaving, 
\begin{align}
\mathcal{F}_\mathfrak{n}^{(L)} &= \sum\nolimits_{a}  \sqrt{\tfrac{2E^\star_\mathfrak{n}}{-\mu^{\prime\star}_0}} \, (\mathbf{w}_0)_a \tfrac{1}{k_a^\star}\,  \mathcal{F}_a \qty(s=E^{\star2}_\mathfrak{n})\,, \nonumber\\
&= \sum\nolimits_a \tilde{r}_{\mathfrak{n},a}(L)\, \mathcal{F}_a\qty(s=E^{\star2}_\mathfrak{n})\,, 
\label{eq:FVcorrec}
\end{align}
where the `finite volume form factor', $\mathcal{F}_\mathfrak{n}^{(L)}$, is defined in terms of the matrix-element on the left-hand-side of Eq.~\ref{eq:LLform} as,
\begin{equation}\label{eq:fv_kin_decomp}
\mel{\mathfrak{n}}{\mathcal{J}^\mu(x=0)}{0}_L = \tfrac{1}{\sqrt{2E_\mathfrak{n}L^3}}\, K^\mu\,  \mathcal{F}_\mathfrak{n}^{(L)} \,,
\end{equation}
where $K^\mu$ is given by the straightforward subduction of $K^\mu_m$ in Eq.~\ref{eq:pw_kin} into the irrep of the finite-volume state $\ket{\mathfrak{n}}_L$, as described in Appendix~\ref{sec:kin_fac_subd}.

Equation~\ref{eq:FVcorrec} provides the relationship between the computed finite-volume form factors, of which there is one per finite volume energy level, $\mathcal{F}_\mathfrak{n}^{(L)}$, and the infinite-volume form factors, $\mathcal{F}_a$, for each kinematically open channel $a$, evaluated at $s=E^{\star2}_\mathfrak{n}$. The vector of correction factors, $\tilde{r}_{\mathfrak{n},a}(L)$, is available to us provided we have determined the relevant scattering amplitudes, $\mathcal{M}(s)$, from the finite-volume spectrum. 

Clearly, solving Eq.~\ref{eq:FVcorrec} energy-by-energy for the $\mathcal{F}_a$ in a coupled-channel situation is not possible, but if the form factors have their energy dependence parameterized, then by considering multiple energy-levels, a system of equations can be set up and a best description of multiple $\mathcal{F}_\mathfrak{n}^{(L)}$ values found, yielding the energy dependence of the form factors.

\section{Lattice technology and correlator computation}\label{sec:latt}

The calculation to be presented in this paper was performed on an $(L/a_s)^3\times (T/a_t)= 24^3\times 256$ lattice with two degenerate light $u,d$ quarks and a heavier $s$ quark. The light quark mass is set to be heavier than the physical value, yielding pions of mass \mbox{284 MeV} and kaons of mass \mbox{519 MeV}. Anisotropy introduced in the lattice action gives rise to a temporal spacing $\xi = 3.455(6)$ times finer than the spatial spacing. Comparing to the physical $\Omega$ baryon mass yields $a_t^{-1} \approx 5.988(17)~\mathrm{GeV}$. In the temporal lattice units used throughout this paper, the pion and kaon masses are $a_t\, m = 0.04735(22),\, 0.08659(14)$ respectively.
This lattice has been used in previous computations of $\pi\pi$ scattering~\cite{Rodas:2023twk,Rodas:2023gma}, $\pi K$ scattering~\cite{Wilson:2019wfr} and the electromagnetic transition $\gamma K \to K\pi$~\cite{Radhakrishnan:2022ubg}. 

\bigskip

The finite-volume spectrum needed to constrain scattering amplitudes is extracted from matrices of two-point correlation functions computed using a basis of operators described below which make use of \emph{smeared} quark fields, and we opt to smear with the \emph{distillation}~\cite{HadronSpectrum:2009krc} approach, using the lowest 162 eigenvectors of the gauge covariant spatial Laplacian. In this way two-point functions factorize into pieces dependent on the particular operator constructions, and pieces which encode the propagation of the quarks, known as perambulators.

In order to access \emph{production} amplitudes for $\gamma \to \pi \pi$ and $\gamma \to K \overline{K}$, two-point correlation functions featuring the \emph{local} vector current operator $\mathcal{J}^\mu(x)$ on the sink timeslice are required, and the quark fields appearing in the vector current should not be smeared. This requires the computation of \emph{generalized perambulators} which feature an insertion of the unsmeared vector current. Those used in the present calculation are a re-used subset of those computed for use in Ref.~\cite{Radhakrishnan:2022ubg}. Only light-quark generalized perambulators are required, since only the isovector component of the electomagnetic current can feature in this case. The correlation functions describing production were computed on an ensemble of 348 configurations.\footnote{These 348 configurations are a subset of an ensemble of 400 configurations used to compute the spectrum.}

\medskip 

An established procedure to obtain the discrete spectrum of finite-volume eigenstates in a lattice calculation proceeds by diagonalizing the matrix whose elements are ${C_{ij}(t) = \langle 0 | \mathcal{O}^{\phantom{\dag}}_i(t) \, \mathcal{O}^\dag_j(0) | 0 \rangle }$, with $\{ \mathcal{O}_i \}_{i=1 \ldots N}$ being a basis of operators having the quantum numbers of the states of interest. In practice the diagonalization is best achieved through solution of a generalized eigenvalue problem,
\begin{equation} \label{eq:gevp}
 \mathbf{C}(t) \, v_\mathfrak{n} = \lambda_\mathfrak{n}(t,t_0) \, \mathbf{C}(t_0) \, v_\mathfrak{n} \, ,
\end{equation}
where the time-dependence of the eigenvalues can be fitted to give the energies $E_\mathfrak{n}$, and where the particular linear combination of basis operators given by the eigenvector weights, $\Omega^\dag_\mathfrak{n} =\sqrt{2E_\mathfrak{n}} e^{-E_\mathfrak{n} t_0 /2}\sum\nolimits_i (v_\mathfrak{n})_i\,  \mathcal{O}^\dag_i$, optimally interpolates the discrete state $|\mathfrak{n}\rangle$, and is approximately orthogonal to other states in the spectrum in the sense that $\mel{\mathfrak{n}}{\Omega^\dagger_\mathfrak{m}(0)}{0} = \sqrt{2E_\mathfrak{n}}\,  \delta_{\mathfrak{n},\mathfrak{m}}$, up to small corrections.

The specific operator basis used in this calculation includes fermion bilinears which resemble $q\bar{q}$ constructions and meson-meson-like operators resembling $\pi\pi$ and $K\overline{K}$ pairs with relative momentum. The set of allowed relative momenta is guided by the energy such pairs would have given no dynamical interactions between them, with inclusion of all operators producing a state below the $\phi \pi$ threshold at $a_t E^\star \sim 0.22$.
The opening of the $\omega \pi$ and $\phi \pi$ channels in the energy region of interest led us to also add operators resembling these pairs, but it was found that upon their addition there was negligible change in the energy of states with large overlap onto $\pi\pi$-like and $K\overline{K}$-like operators, and as such we will present results excluding $\omega \pi, \phi \pi$ operators from the basis.\footnote{See Appendix~\ref{app:spectrum} for a presentation of this, and Refs.~\cite{Woss:2018irj, Woss:2019hse} for some discussion of the construction of pseudoscalar-vector meson pair operators.}

\begin{figure}
\includegraphics[width=0.55\columnwidth]{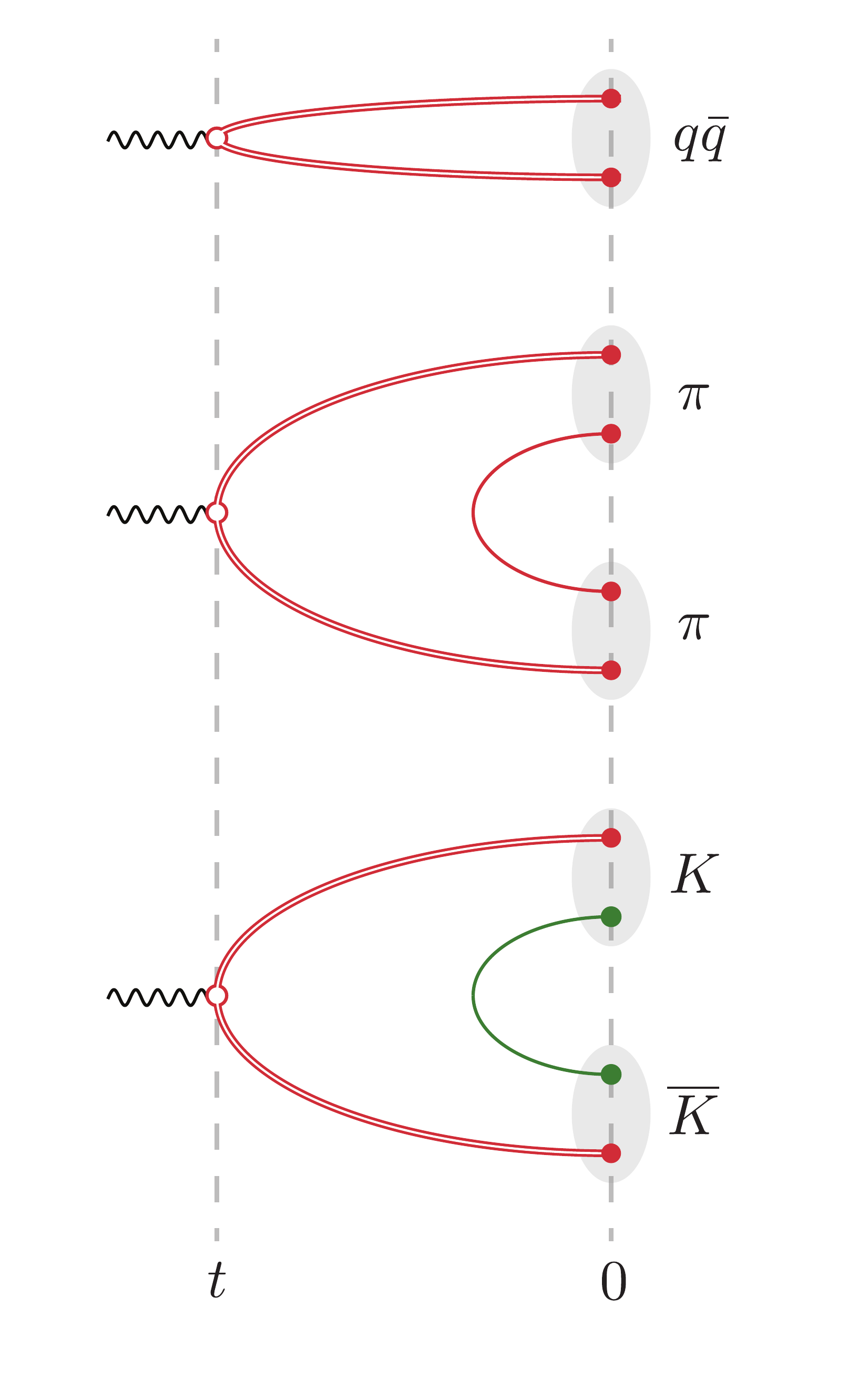}
\caption{\label{fig:wicks}Wick contraction diagrams required to compute $\langle 0 | \mathcal{J}(t) \, \Omega^\dag_\mathfrak{n}(0) | 0 \rangle$, where single red and green lines represent light-quark and strange quark perambulators respectively, and where the double line represents the light-quark generalized perambulator housing the local current insertion. }
\end{figure}

The optimized operators found by solving the generalized eigenvalue problem described above are used in the subsequent computation of correlation functions featuring the vector current, $\langle 0 | \mathcal{J}(t) \, \Omega^\dag_\mathfrak{n}(0) | 0 \rangle$, and these correlation functions are expected to be dominated by the contribution of the specific finite-volume state $|\mathfrak{n}\rangle$, with other states, including those of lower energy, being suppressed~\cite{Shultz:2015pfa}. The extent to which this is true in practical calculation will be explored in Section~\ref{sec:timefits}. In order to compute these correlation functions featuring the unsmeared vector current, diagrams of the type shown in Figure~\ref{fig:wicks} must be computed, featuring the generalized perambulators with current insertion at time $t$.

\section{Energy spectrum and partial wave scattering amplitudes} \label{sec:spectrum}

\begin{figure*}
\includegraphics[width=\textwidth]{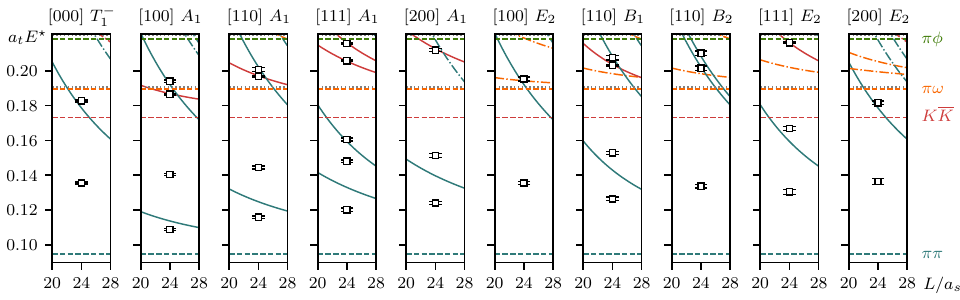}
\caption{\label{fig:full_spec}Finite volume spectra obtained for all irreps considered in this analysis. 
Dashed lines show meson-meson thresholds, while solid lines show non-interacting meson-meson energies as a function of $L$, color-coded according to the meson pair (dot-dashed indicating non-interacting energies where the associated operator was not included in the variational basis).
Grey dotted lines indicate thresholds for higher-multiplicity scattering ($\pi\pi\pi\pi$, $\pi\pi\eta$, $K\overline{K}\pi$). 
}
\end{figure*}

Matrices of correlation functions using the basis of operators described in the previous section were computed for several irreps at rest and in moving frames. Variational analysis of these using Eq.~\ref{eq:gevp} leads to spectra that densely span the elastic scattering region and the lower part of the coupled-channel region.\footnote{See also Ref.~\cite{Rodas:2023gma} for a similar extraction on the same lattices, focusing on the elastic region.}
Together they provide a high degree of constraint on the $P$-wave partial-wave scattering amplitudes over the region of interest, and the corresponding optimized operators will be used to determine the production amplitudes at each of the dense set of discrete energies.

Figure~\ref{fig:full_spec} presents the extracted spectra, with the levels below $K\overline{K}$ threshold showing clear departures from non-interacting $\pi\pi$ energies, as well as a counting indicating an `additional' level. An avoided-level-crossing is apparent whenever a non-interacting level lies near ${a_t E^\star \sim 0.13}$. These observations strongly suggest the presence of a narrow resonance as confirmed below through determination of scattering amplitudes. In the coupled-channel region above $K\overline{K}$ threshold, the extracted energy levels are seen to lie very close to non-interacting values, suggesting that here the interactions are weak and that the scattering system is free of additional resonances below ${a_t E^\star \sim 0.23}$.

\vspace*{-6mm}

\subsection{Elastic \texorpdfstring{$\pi\pi$}{ππ} scattering}\label{sec:elas_scat}

\begin{figure}
\includegraphics[width =\columnwidth]{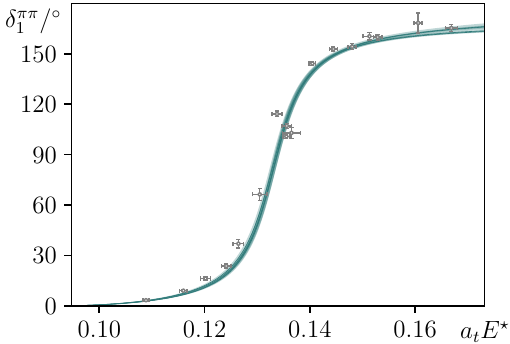}
\caption{Elastic $\pi\pi$ $P$-wave phase-shift constrained by energy levels below $K\overline{K}$ threshold in Fig.~\ref{fig:full_spec}. Bands shows parameterizations including the \emph{reference elastic} parameterization of Eq.\ref{eq:elastic_ref}.}
\label{fig:ps_el}
\end{figure}

In the $\pi\pi$ elastic region below $K\overline{K}$ threshold, if the effect of $J=3$ and higher spin amplitudes can be neglected, the Lüscher quantization condition of Eq.~\eqref{eq:Luscherqc} becomes an algebraic relation for the elastic $\pi\pi$ scattering amplitude, with each discrete energy level yielding a value of the $J^P=1^-$ phase-shift at that energy. Alternatively, all energy levels can be described together by parameterizing the energy-dependence of the elastic scattering amplitude, and minimizing a $\chi^2$ as explained in Ref.~\cite{Wilson:2014cna}. These two approaches are presented in Figure~\ref{fig:ps_el}, where four parameterization choices are shown, each of which prove capable of describing 17 energy levels below $a_t E^\star = 0.173$ with a $\chi^2/N_\text{dof}\approx 1.2$ (see Appendix~\ref{sec:scat_variation}). As can be seen the result is not sensitive to the details of the parameterization, and in the remainder of this manuscript we will focus on a \emph{reference elastic} amplitude constructed from a $K$-matrix built from a pole and a constant, and a dispersively improved `Chew-Mandelstam' phase-space\footnote{See e.g. Appendix B of Ref.~\ref{Wilson:2014cna}} subtracted at $s=m^2$, the $K$-matrix pole location,
\begin{align}
  \mathcal{M}(s) &= \frac{16\pi}{ \tfrac{1}{(2k^\star)^2}\mathcal{K}^{-1}(s) -I_{\text{CM}}(s)} \,, \nonumber \\
  \mathcal{K}(s) &= \frac{g^2}{m^2 - s} +\gamma \, ,\label{eq:K_mat_el}
\end{align}
where the best fit to the lattice energy levels is with parameter values,
\begin{center}
	\begin{tabular}{rll}
	$m =$                         & $0.1335\, (5) \cdot a_t^{-1}$   &
	\multirow{3}{*}{ $\begin{bmatrix*}[r] 1 &  -0.3 & -0.3  \\
					                    & 1 & 0.7  \\
					                    &   & 1    \end{bmatrix*}$ } \\
	$g = $                  & $0.445\,(10)$   & \\
	$\gamma = $             & $(3.4\pm 2.2) \cdot a_t^{2}$   & \\[1.3ex]
	\multicolumn{3}{r}{$\chi^2/N_{\text{dof}}=\frac{17.0}{17-3}=1.21$\, .}
	\end{tabular}
\end{center}\vspace{-1.0cm} \begin{equation}\label{eq:elastic_ref}\end{equation}

The behavior observed in Fig.~\ref{fig:ps_el} is clearly that of a narrow resonance, and indeed the \emph{reference elastic} amplitude is found to feature a pole on the unphysical Riemann sheet at
\begin{equation*}
a_t\sqrt{s_R} = 
0.1328(5) -\tfrac{i}{2}0.0096(5) \, ,
\end{equation*}
with a coupling to $\pi\pi$ defined at the pole, ${\mathcal{M}(s \sim s_R) \sim 16\pi \, \frac{c_{\pi\pi}^2}{s_R-s}}$,
of value,
\begin{equation*}
a_t\,c_{\pi\pi} = 0.0426(11)\,e^{-i\pi\cdot 0.047(3)}\, .
\end{equation*}

\subsection{Coupled \texorpdfstring{$\pi\pi$/$K\overline{K}$}{ππ/KK} scattering}\label{sec:cc_scat}

Above $K\overline{K}$ threshold, each energy level is in principle sensitive to all elements of the coupled-channel scattering matrix, and as such we must proceed using parameterizations of this matrix. The structure of Eq.~\ref{eq:Luscherqc} is such that only parameterizations consistent with unitarity will generate solutions. Use of $K$-matrix forms ensures unitarity, and parameterizations in which the $K$-matrix is constructed from poles plus polynomials in $s$ have been successful in past attempts to describe lattice spectra, and in particular in Ref.~\cite{Wilson:2015dqa} where the $\pi\pi, K\overline{K}$ $P$-wave system was considered at a lighter pion mass.

\begin{figure}
\includegraphics{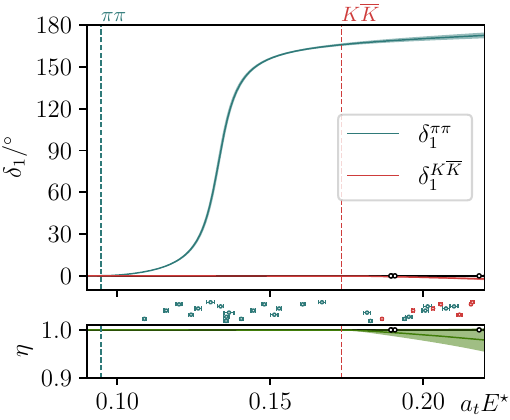}
\caption{\label{fig:ps_cc}
Phase-shifts, $\delta_1^a$,  and inelasticity, $\eta$, for coupled-channel $\pi\pi, K\overline{K}$ scattering corresponding to the amplitude given in Eq.~\ref{eq:cc_ref}. The constraining energy level locations are shown between the panels, with levels lying close to $K\overline{K}$ non-interacting energies colored in red. White circles show the thresholds for $\pi\pi\pi\pi$, $\omega \pi$ and $\phi \pi$ production, in that order.}
\end{figure}

A simple form, which we will refer to as the \emph{reference coupled-channel} amplitude, capable of describing the entire spectrum, is given by,
\begin{align}
  \big[\mathcal{M}^{-1}\big]_{ab}(s) &= \tfrac{1}{16\pi} \left( \tfrac{1}{2k_a^\star} \big[\mathcal{K}^{-1}(s)\big]_{ab} \tfrac{1}{2k_b^\star}- \delta_{ab}\,  I_{\text{CM},a}(s)  \right)\, , \nonumber \\
  \mathcal{K}_{ab}(s) &= \frac{g_a \, g_b}{m^2 - s} +\gamma_{ab} \, ,\label{eq:cc_ref_kmatrix}
\end{align}
(where a Chew-Mandelstam phase-space is used) with parameter values,
\footnotesize
\begin{center}
	\begin{tabular}{rll}
	$m =$                         & $0.1338\, (5) \cdot a_t^{-1}$   &
	\multirow{6}{*}{ $\begin{bmatrix*}[r] 1 & -0.2 & 0.0  & -0.2 & 0.2  & -0.1  \\
					                        & 1    & -0.4 & 0.6  & -0.4 & -0.4  \\
					                        &      & 1    & -0.2 & 0.8  & 0.9  \\
					                        &      &      & 1    & -0.2 & -0.1  \\
					                        &      &      &      & 1    & 0.8  \\
					                        &      &      &      &      & 1  \\    
					                        \end{bmatrix*}$ } \\
	$g_{\pi\pi}   = $                  & $0.441\,(9)$   & \\
	$g_{K\overline{K}} = $                  & $0.17\,(30)$   & \\
	$\gamma_{\pi\pi, \pi\pi} = $       & $(2.9 \pm 0.9) \cdot a_t^{2}$   & \\
	$\gamma_{\pi\pi, K\overline{K}} = $     & $-(2.4 \pm 5.0) \cdot a_t^{2}$   & \\
	$\gamma_{K\overline{K}, K\overline{K}} = $   & $-(2.2 \pm 4.0) \cdot a_t^{2}$   & \\[1.3ex]
	\multicolumn{3}{l}{\quad\quad\quad\quad $\chi^2/N_{\text{dof}}=\frac{28.7}{32-6}=1.10$\,.}
	\end{tabular}
\end{center}\vspace{-0.9cm} 
\normalsize
\begin{equation}\label{eq:cc_ref}\end{equation}
This amplitude is presented in Fig.~\ref{fig:ps_cc} in terms of the $\pi\pi$ and $K\overline{K}$ phase-shifts and the inelasticity, whose departure from value 1 indicates channel coupling\footnote{The definition of these quantities in terms of the scattering amplitude is given by Eq.~(8) of Ref.~\cite{Wilson:2015dqa}.}. As was suggested by the spectra shown in Figure~\ref{fig:full_spec}, the amplitudes indicate no strong scattering above the $K\overline{K}$ threshold, and only mild evidence for $\pi\pi, K\overline{K}$ channel coupling.
The $\rho$ resonance is clearly present in the $\pi\pi$ phase-shift, and because we use a coupled-channel parameterization at all energies, in principle the $\rho$--pole will have a coupling to the $K\overline{K}$ channel. Such a resonance coupling can be rigorously defined by factorizing the amplitude at the location of the complex resonance pole on the unphysical Riemann sheet,
\begin{equation}
\mathcal{M}_{ab}(s \sim s_R) \sim 16\pi \, \frac{c_a \, c_b}{s_R-s} \, ,
\end{equation}
however, because the $\rho$ lies well below the $K\overline{K}$ threshold, the coupling to $K\overline{K}$ will not be very well constrained as the impact of the $K\overline{K}$ components of $\mathcal{M}$ is exponentially suppressed in Eq.~\ref{eq:Luscherqc} below $K\overline{K}$ threshold. The amplitude presented above has a pole\footnote{On the sheet closest to physical scattering below $K\overline{K}$ threshold where the imaginary parts of the $\pi\pi$ and $K\overline{K}$ momenta are negative and positive, respectively -- sheet II in the usual nomenclature.} at
\begin{equation}\label{eq:sr_ref_cc}
a_t\sqrt{s_R} = 
0.1331(4) -\tfrac{i}{2}0.0095(4) \, ,
\end{equation}
with channel couplings,
\begin{align}
a_t\,c_{\pi\pi}        &= 0.0424(8)\, e^{-i \pi \cdot 0.047(2)}\,, \nonumber \\
a_t\,c_{K\overline{K}} &= 0.019(33) \, e^{i \pi \cdot 0.47(5)} \, , \label{eq:coup_ref_cc}
\end{align}
which are observed to be close to real valued for the kinematically open $\pi\pi$ channel, and close to imaginary valued for the kinematically closed $K\overline{K}$ channel.

The pole location in this coupled-channel analysis is seen to be compatible with the one found in the previous elastic analysis, as expected for a resonance lying well below the $K\overline{K}$ threshold. Further, also as expected, the $K\overline{K}$ coupling is not precisely determined. Owing to the $P$-wave nature of the resonance, the (dimensionless) values of the couplings with the barrier factor divided out are also relevant,
\begin{align*}
  \sqrt{16\pi}\, \abs{\hat{c}_{\pi\pi}} &\equiv \sqrt{16\pi}\left| \frac{ c_{\pi\pi} }{ k^\star_{\pi\pi}(s_R) } \right| = 6.41 \pm 0.13  \,, \nonumber \\
  \sqrt{16\pi}\, \abs{\hat{c}_{K\overline{K}}} &\equiv  \sqrt{16\pi}\left| \frac{ c_{K\overline{K}} }{ k^\star_{K\overline{K}}(s_R) } \right| = 2.4 \pm 4.0 \, .
\end{align*}

As well as this \emph{reference coupled-channel} amplitude, parameterization variations listed in Appendix~\ref{sec:scat_variation} also prove capable of describing the spectra, generating plots which closely resemble Fig.~\ref{fig:ps_cc} and which have a $\rho$ pole and couplings broadly compatible with those given above.

\pagebreak
\section{Lattice matrix elements}\label{sec:timefits}

To obtain \emph{production} amplitudes, the required finite-volume matrix elements are extracted from two-point correlation functions, $\langle 0 | \mathcal{J}(t) \, \Omega^\dag_\mathfrak{n}(0) | 0 \rangle$, which feature the electromagnetic current. This current can be decomposed into components of definite isospin,
\begin{equation}\label{eq:curr_renorm}
\mathcal{J}= Z_V^l\tfrac{1}{\sqrt{2}}\left(\mathcal{J}_{\rho, {\text{lat}}}+\tfrac{1}{3}\mathcal{J}_{\omega_l,\text{lat}} \right) + Z_V^s \left( -\tfrac{1}{3} \mathcal{J}_{\omega_s, \mathrm{lat}} \right)\,,
\end{equation}
where the isospin-basis currents are
\begin{equation*}
  \mathcal{J}_{\rho} \equiv \tfrac{1}{\sqrt{2}} \big( \bar{u} \Gamma u - \bar{d} \Gamma d \big),
  \mathcal{J}_{\omega_l} \equiv \tfrac{1}{\sqrt{2}} \big( \bar{u} \Gamma u + \bar{d} \Gamma d \big),
  \mathcal{J}_{\omega_s} \equiv  \bar{s} \Gamma s \, ,
\end{equation*}
and the spatially-directed vector current whose improvement at $\mathcal{O}(a)$ is consistent with the anisotropic Clover quark action is~\cite{Shultz:2015pfa},
\begin{equation}\label{eq:curr_impro}
  \bar{q} \Gamma q = \bar{q} \gamma^k q + \tfrac{1}{4} (1- \xi)\, a_t \partial_4 \big(\bar{q} \sigma^{4k} q \big) \, .
\end{equation}

In the case where production of $\pi\pi$ is considered, only the isovector component $\mathcal{J}_{\rho}$ appears, and the multiplicative renormalization factor, $Z_V^l$, is determined non-perturbatively using the pion form factor at zero virtuality extracted from three-point correlation functions, $\langle 0 | \Omega_\pi(\Delta t) \, \mathcal{J}(t) \, \Omega^\dag_\pi(0) | 0 \rangle$, as described in Ref.~\cite{Radhakrishnan:2022ubg}, and Appendix~\ref{sec:pi_3pt}. 

\smallskip

We compute two-point current correlation functions, $\langle 0 | \mathcal{J}(t) \, \Omega^\dag_\mathfrak{n}(0) | 0 \rangle$, for each energy level in Figure~\ref{fig:full_spec} using the corresponding optimized operator determined by solving the relevant generalized eigenvalue problem.
The current operator is subduced into the irrep $\Lambda$ and projected to spatial momentum $\mathbf{P}$ of the state $\mathfrak{n}$, and the correlation function is computed for all $0\leq t/a_t<32$. The use of an optimized operator for state $\mathfrak{n}$ should ensure a leading time-dependence of
\begin{equation*}  
\langle 0 | \mathcal{J}(t)^{}\, \Omega^\dag_\mathfrak{n}(0) |0\rangle
=  e^{-E_\mathfrak{n}t} \, \sqrt{2E_\mathfrak{n}}  \, L^3\mel{0}{\mathcal{J}(x=0)}{\mathfrak{n}}_L +\dots\,,
\end{equation*}
where the subleading time-dependence is expected to arise from overlap with high-lying states. Hence, fitting $ e^{E_\mathfrak{n} t}\cdot \langle 0 | \mathcal{J}(t)^{}\, \Omega^\dag_\mathfrak{n}(0) |0\rangle$ to a constant (at late times) or a constant plus an exponential modelling excited states (when including earlier times) would yield the desired matrix element in the fitted constant.

In practice, timeslice-to-timeslice data correlation is reduced if the following ratio is instead considered:
\begin{equation*} 
\frac{\langle 0 | \mathcal{J}(t)^{}\, \Omega^\dag_\mathfrak{n}(0) |0\rangle}
{\langle 0 | \Omega_\mathfrak{n}(t) \, \Omega^\dag_\mathfrak{n}(0) |0\rangle}
=  \frac{L^3}{\sqrt{2E_\mathfrak{n}}}\mel{0}{\mathcal{J}(x=0)}{\mathfrak{n}}_L +\dots\,,
\end{equation*}
which can be fitted to a constant or a constant plus an exponential.
For convenience, we define the dimensionless ratio,
\begin{equation} \label{eq:Rdef}
R_\mathfrak{n}(t) \equiv \frac{\sqrt{2 E_\mathfrak{n}}}{K(\Lambda)}\frac{\langle 0 | \mathcal{J}(t)^{}\, \Omega^\dag_\mathfrak{n}(0) |0\rangle}
{\langle 0 | \Omega_\mathfrak{n}(t) \, \Omega^\dag_\mathfrak{n}(0) |0\rangle} = \sqrt{\tfrac{L^3}{2 E_\mathfrak{n}}} \mathcal{F}_\mathfrak{n}^{(L)} +\dots\,,
\end{equation}
where we have also divided out the kinematic factor $K(\Lambda)$, which corresponds to the factor $K^\mu$ of Eq.~\eqref{eq:fv_kin_decomp} once the current operator has been subduced into irrep $\Lambda$ (see Tab.~\ref{tab:kin_fac} in App.~\ref{sec:kin_fac_subd}).

\begin{figure}
\includegraphics{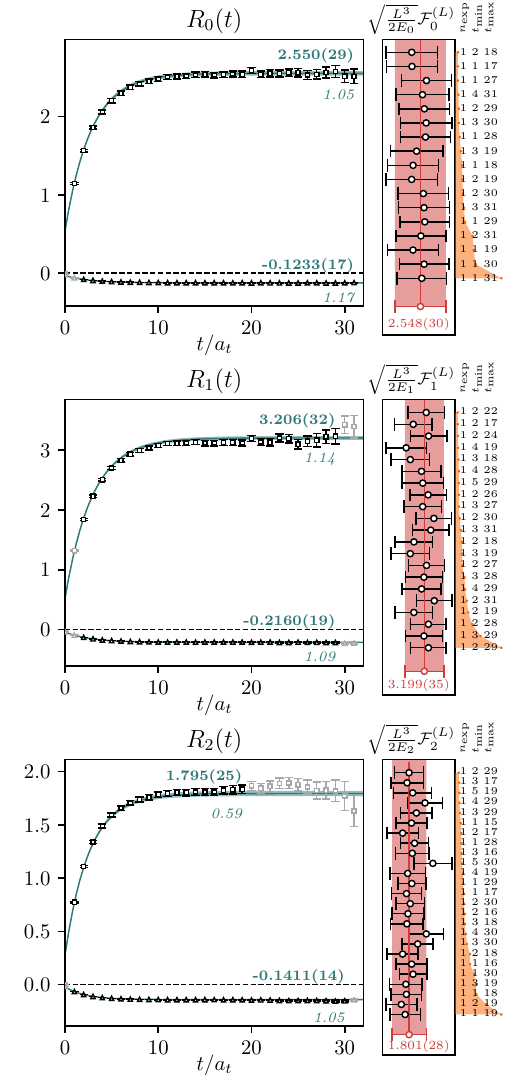}
\caption{\label{fig:mel_td_fit}Time dependence of $R_\mathfrak{n}(t)$ defined in Eq.~\ref{eq:Rdef} for the three lowest discrete energy levels in the ${[111]\,A_1}$ irrep. Squares/triangles show data for the unimproved/improvement term currents (first and second terms in Eq.~\eqref{eq:curr_impro}). Curves show the timeslice description having largest AIC weight, with the constant fit value and the $\chi^2/N_{\text{dof}}$ of the fit also shown. Variations in the constant fit value for the unimproved current over different timeslice fit windows are shown in the right column, together with the AIC weight (in orange), and the ``model average'' (in red).}
\end{figure}

The timeslice behavior of $R_\mathfrak{n}(t)$ for the three lowest lying states of the [111] $A_1$ irrep is shown in Fig.~\ref{fig:mel_td_fit}. The observed relaxation to a constant value for ${t \gtrsim t_0 = 10 \, a_t}$, even for states above the ground state, is a result of making use of optimized operators.
We show separately the contributions of the two terms in Eq.~\eqref{eq:curr_impro}, where the $\mathcal{O}(a)$ improvement term is observed to impact at the level of $5-10\%$ of the leading term. For the anisotropic lattice considered in this work, the improvement enters proportional to the energy difference between the initial and final states of the matrix element, which is larger in magnitude in the current case of production than in the case of three-point functions used to extract the spacelike pion form factor, see App.~\ref{sec:pi_3pt}.

The $R_\mathfrak{n}(t)$ timeslice data can be described using constant or constant plus exponential forms fitted over various windows. We show the result of using a version~\cite{Jay:2020jkz} of the Akaike Information Criterion (AIC) which assigns a weight to each fit depending on its correlated $\chi^2$ and the number of degrees of freedom, allowing us to average over a range of fits. In practice we observe that the `model average' is dominated by only four or five fits which have large weight and the averaged error is dominated by the statistical error on these. The model average in each case is used for all subsequent analysis in this paper.

\begin{figure*}
\includegraphics{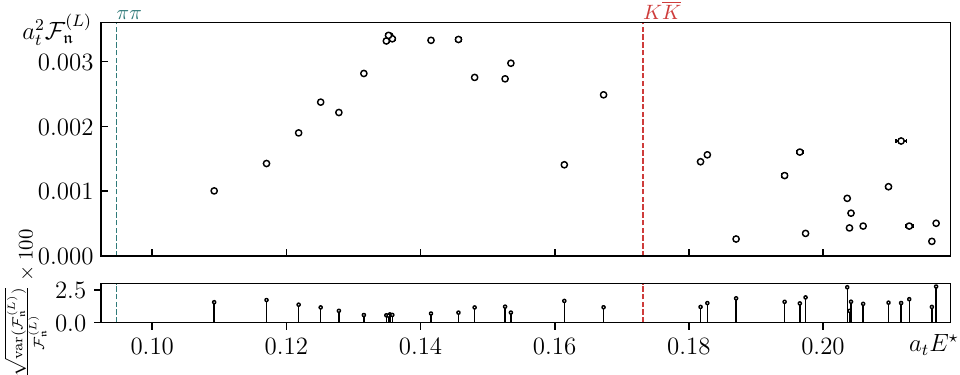}
\caption{\label{fig:FL_data}Finite volume form factor in the $\pi\pi$ elastic region and above the $K\overline{K}$ threshold. The fractional statistical uncertainties for the form factor values are shown in the bottom panel.}
\end{figure*}

Levels found lying near $K\overline{K}$ non-interacting energies (for example the fourth state in $[111]\, A_1$) demand a slightly different timeslice analysis.
The heavily suppressed value of $\mathcal{F}^{(L)}_\mathfrak{n}$ for these states relative to the large $\mathcal{F}^{(L)}_\mathfrak{m}$ for lower-lying states near the resonance energy, is such that even though the optimized operator $\Omega_\mathfrak{n}$ has only a tiny overlap onto lower-lying states $|\mathfrak{m}\rangle$, the combination $\mathcal{F}_\mathfrak{m}^{(L)} \cdot  \mel{\mathfrak{m}}{\Omega_{\mathfrak{n}}^\dagger(0)}{0}$ can be significant. This gives rise to a `subleading' contribution to $R_\mathfrak{n}(t)$ with exponentially \emph{growing} time-dependence,
\begin{equation*}
R_\mathfrak{n}(t) = \dots + \sum_{\mathfrak{m} < \mathfrak{n}} \varepsilon_{\mathfrak{m},\mathfrak{n}} \, e^{(E_\mathfrak{n}-E_\mathfrak{m})t} + \dots \, ,
\end{equation*}
where $\varepsilon_{\mathfrak{m},\mathfrak{n}} \propto \mathcal{F}_\mathfrak{m}^{(L)} \cdot  \mel{\mathfrak{m}}{\Omega_{\mathfrak{n}}^\dagger(0)}{0}$. Approaches to handle these cases are presented in Appendix~\ref{sec:late_t_poll}.

\medskip 

The resulting values of $\mathcal{F}_\mathfrak{n}^{(L)}$ from timeslice fits for energy levels across all irreps considered are summarized in Fig.~\ref{fig:FL_data}. The data is visibly enhanced near the resonance energy ${a_t E^\star \sim 0.135}$, while for higher energies it does not appear to have any simple energy dependence, with a strong dependence on the irrep of state $\mathfrak{n}$ being observed even for very similar values of $a_t E^\star_{\mathfrak{n}}$.
This is as expected given the need for finite-volume corrections, which will be addressed in the next section.

\section{Finite volume correction factor} \label{sec:LL}

The finite-volume correction factors, $\tilde{r}_{\mathfrak{n},a}(L)$, defined in Eq.~\ref{eq:FVcorrec}, are obtained making use of parameterized scattering matrices, $\mathcal{M}(s)$, that describe the finite-volume spectra as described in Section~\ref{sec:spectrum}. The matrix $\mathcal{M}^{-1}+F$ is eigen-decomposed at the energy where one of its eigenvalues crosses zero, which approximates within the scattering `model' the lattice QCD computed energy, $E_\mathfrak{n}(L)$. At this point the eigenvector, $\mathbf{w}_0$, corresponding to the zero eigenvalue is obtained, and this is used in the computation of the required eigenvalue slope in a finite-difference,
\begin{align}   \label{eq:slope}
  \mu^{\prime\star}_0 = \frac{1}{2\, \Delta E} \,
\mathbf{w}_0^\intercal \cdot \bigg(
\big[ &\mathcal{M}^{-1}+F \big]_{E_\mathfrak{n} + \Delta E}  \nonumber\\
&- \big[\mathcal{M}^{-1}+F\big]_{E_\mathfrak{n} - \Delta E} 
\bigg) \cdot \mathbf{w}_0\, ,
\end{align}
which enters into
\begin{equation} \label{eq:rtilde_expl}
  \tilde{r}_{\mathfrak{n},a}(L)=\sqrt{\frac{2E^\star_\mathfrak{n}}{-\mu^{\prime\star}_0}}\frac{\mathbf{w}_{0,a}}{k_a^\star}\,.
\end{equation}
Uncertainties are propagated into this quantity using jackknife via the ensemble of scattering matrix parameter values.

As shown in Appendix B of Ref.~\cite{Radhakrishnan:2022ubg}, in the case that $\mathcal{M}(s)$ houses a \emph{narrow} resonance, for energies near to the resonance mass, the finite-volume correction factor becomes volume-independent and has elements proportional to the coupling of the resonance to each channel, $a$,
\begin{equation*}
  \tilde{r}_{\mathfrak{n},a}(L) \approx \sqrt{16\pi}\;  \frac{c_{a}}{k^\star_a} + \mathcal{O}\left(\tfrac{\Gamma_R}{m_R} \right)\,.
\end{equation*}
This reflects the dominance in the scattering at these energies of a spatially localized state whose wavefunction does not sample the boundary of the finite-volume. 

At energies where $\mathcal{M}$ describes \emph{weak} scattering and the solutions to Eq.~\ref{eq:Luscherqc} lie close to non-interacting meson-pair energies, the value of $\tilde{r}_{\mathfrak{n},a}(L)$ is set largely by properties of the geometric matrix $F(E, L)$ related to how the meson-pair relative momentum directions subduce into the irrep under consideration. Between these two extremes the finite-volume correction factor is sensitive to both $\mathcal{M}$ and $F(E,L)$.

\subsection{Elastic finite volume correction}

\begin{figure}[b]
\includegraphics{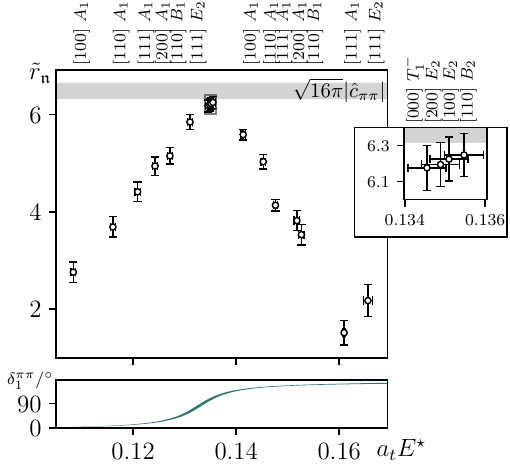}
\caption{\label{fig:rtilde}Elastic $\pi\pi$ finite-volume factors, $\tilde{r}_\mathfrak{n}$, computed using the \emph{reference elastic} scattering amplitude parameterization.}
\end{figure}

Below $K\overline{K}$ threshold, where only elastic $\pi\pi$ scattering is relevant, $\tilde{r}_\mathfrak{n}$ can be considered to be just a single number correcting the normalization of the matrix element for each finite-volume energy level. In Fig.~\ref{fig:rtilde} we present values of $\tilde{r}_\mathfrak{n}$ for the 17 energy levels below $K\overline{K}$ threshold, computed using the \emph{reference elastic} amplitude. 

The very similar values observed for four energy levels in different irreps close to $a_t E^\star = 0.135$ are explained by the presence of the narrow $\rho$ resonance, with the values of $\tilde{r}_\mathfrak{n}$ being seen to be almost equal to the value of $\sqrt{16\pi} |\hat{c}_{\pi\pi}|$ for this amplitude.

Differing values of $\tilde{r}_\mathfrak{n}$ are observed for the two highest levels presented, each with $\mathbf{P}=[111]$, lying close to the $\pi_{110}\pi_{100}$ non-interacting energy at $a_tE^\star_{\text{non-int}} \approx 0.16$. These values reflect the difference between a nearly non-interacting helicity 0 $\pi\pi$ state subduced into the $A_1$ irrep, and helicity $\pm1$ states subduced into $E_2$, and the same pattern of magnitudes can be observed in the lattice QCD computed matrix elements for the corresponding states in Fig.~\ref{fig:FL_data}.

\subsection{Coupled channel finite volume correction}\label{sec:fv_correc_cc}

In the coupled-channel case, the object in parentheses in Eq.~\ref{eq:slope} is a matrix, with the projection into the zero-crossing eigenvalue achieved with the vector $\mathbf{w}_0$. Above the $K\overline{K}$ threshold, the matrix is real and symmetric, and the components of $\mathbf{w}_0$ are real. Above $\pi\pi$ threshold, but below $K\overline{K}$ threshold this is not the case, and the presence of $\ell=1$ angular momentum barrier factors in $\mathcal{M}$ and $F$ causes $(\mathbf{w}_0)_a \propto k_a^\star$ such that kinematically closed channels become imaginary components of the vector. In the computation of $\tilde{r}_\mathfrak{n}$, this phase is cancelled by the explicit factor of $1/k_a^\star$ in Eq.~\ref{eq:rtilde_expl}, yielding a real valued correction vector (see also the discussion in Appendix~A of Ref.~\cite{Briceno:2021xlc}).

\begin{figure}
\includegraphics[width=\columnwidth]{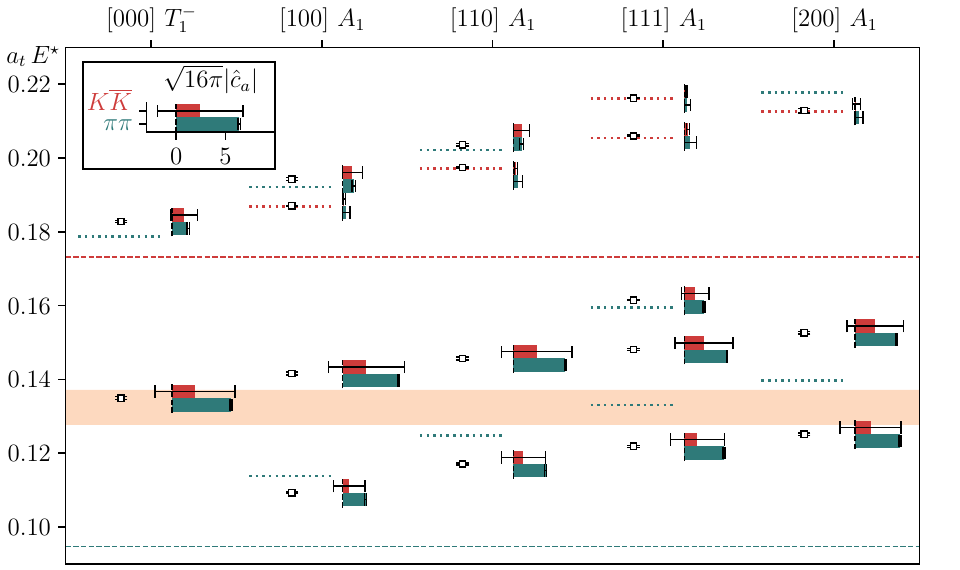}
\includegraphics[width=\columnwidth]{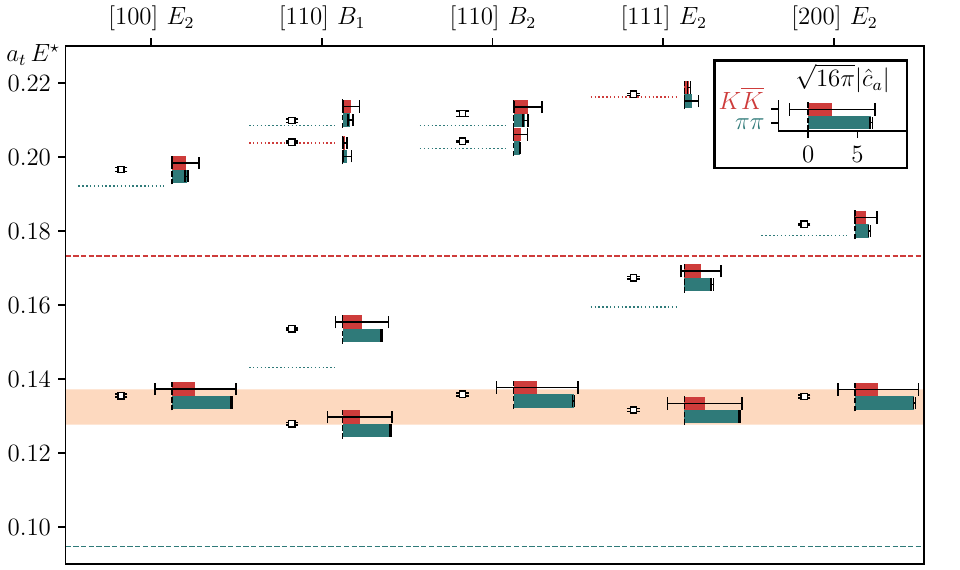}
\caption{\label{fig:rtilde_cc}
Histograms for each energy level show the magnitudes of finite-volume factors, $\big|\tilde{r}_{\mathfrak{n},a} \big|$, for $a=\pi\pi$ (blue bar) and $a=K\overline{K}$ (red bar) computed using the \emph{reference coupled-channel} scattering amplitude. Non-interacting $\pi\pi$ (blue) and $K\overline{K}$(red) energies are indicated by the horizontal dotted lines.}
\end{figure}

Explicit values of $\tilde{r}_{\mathfrak{n},a = \pi\pi,K\overline{K}}$ are shown in Fig.~\ref{fig:rtilde_cc}, computed using the \emph{reference coupled-channel} amplitude parameterization which successfully described the finite-volume spectra presented in Section~\ref{sec:spectrum}.
Levels below $K\overline{K}$ threshold have values of $\tilde{r}_{\mathfrak{n},\pi\pi}$ in close agreement with the values computed using the \emph{reference elastic} amplitude, as presented in Fig~\ref{fig:rtilde}. The corresponding $K\overline{K}$ components are observed to be quite uncertain, as expected given the lack of constraint on $K\overline{K}$ well below its kinematic threshold.\footnote{See Appendix A1 of Ref.~\cite{Briceno:2021xlc} for a discussion of closed channels within this formalism.}
Above $K\overline{K}$ threshold, the $\pi\pi, K\overline{K}$ components are of similar magnitude -- that the $K\overline{K}$ components do not obviously dominate for levels lying close to $K\overline{K}$ non-interacting energies is a result of having already extracted a factor of the scattering matrix, $\mathcal{M}$, from the quantity being finite-volume corrected, see Eq.~\ref{eq:MRM}.

These values of $\tilde{r}_{\mathfrak{n}}$, combined with the lattice QCD computed finite-volume form factors in Fig~\ref{fig:FL_data}, will yield our infinite-volume production form factors.

\section{Timelike pion form factor}\label{sec:ff_tl_el}

\begin{figure}[b]
\includegraphics{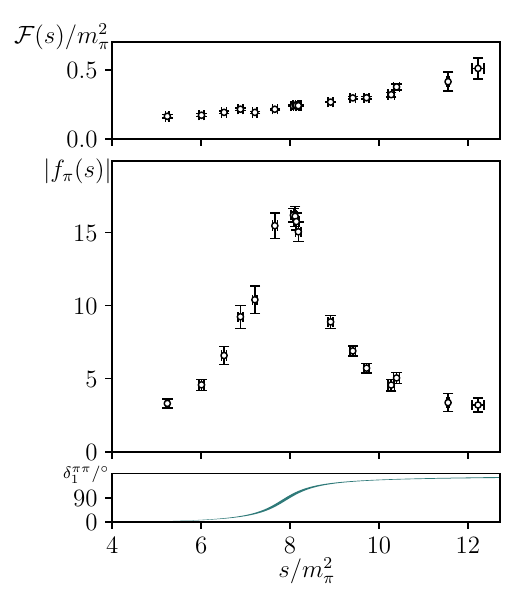}
\caption{\label{fig:f_abs_el} Elastic infinite-volume pion form-factor in the timelike region determined from Eqns.~\ref{eq:Fel_FV_corr}, \ref{eq:MF}. Bottom panel shows the phase-shift for the \emph{reference elastic} scattering amplitude, which by Watson's theorem is also the phase of the production amplitude. 
}
\end{figure}

The finite-volume form factor values extracted from lattice QCD computed correlation functions were presented in Figure~\ref{fig:FL_data}. Focussing initially on those at energies lying below the $K\overline{K}$ threshold, we can correct with the factor $\tilde{r}_\mathfrak{n}$ presented in Figure~\ref{fig:rtilde} to yield discrete values of 
\begin{equation}\label{eq:Fel_FV_corr}
\mathcal{F}\qty(s= E^{\star 2}_\mathfrak{n}) = \frac{1}{\tilde{r}_\mathfrak{n}(L)}\mathcal{F}^{(L)}_\mathfrak{n} \,,
\end{equation}
as shown in the top panel of Fig.~\ref{fig:f_abs_el}.
This data is clearly consistent with being describable by a smooth function of scattering energy, as demanded by unitarity. When weighted with the \emph{reference elastic} scattering amplitude, this data gives the pion form factor in the elastic timelike region (c.f. Eq.~\ref{eq:fNovD}),
\begin{equation}\label{eq:MF}
  f_\pi(s) = \frac{1}{k^{\star \, 2}_{\pi\pi}} \mathcal{M}(s) \, \mathcal{F}(s) \, , 
\end{equation}
which we plot in the middle panel of Figure~\ref{fig:f_abs_el}. The form factor energy dependence is clearly dominated by the presence of the narrow $\rho$ resonance in $\mathcal{M}(s)$.

In order to accurately determine the uncertainty on the function $\mathcal{F}(s)$, we must propagate the correlated errors on the finite-volume correction factors, $\tilde{r}_\mathfrak{n}(L)$, and this poses an implementation challenge. As seen in Figure~\ref{fig:FL_data}, the finite-volume form factor data has tiny fractional errors, such that the dominant uncertainty in Eq.~\ref{eq:FVcorrec} comes from $\tilde{r}_{\mathfrak{n}}(L)$ (see Figures~\ref{fig:rtilde}, \ref{fig:rtilde_cc}), and in practice, the determined values of $\tilde{r}_{\mathfrak{n}}(L)$ for different energy levels, $\mathfrak{n}$, are highly correlated with each other, owing to being generated from scattering amplitude parameterizations featuring a relatively small number of free parameters. 
Naively propagating such a high degree of correlation into the $\chi^2$ minimization which will determine the parameters in a parameterization of $\mathcal{F}(s)$ leads to results which do not reflect the actual level of precision on the original energy level and matrix-element data.

Our approach is to resample the ensembles of $\tilde{r}_{\mathfrak{n}}(L)$, maintaining their computed mean and variance (as shown in Section~\ref{sec:LL}), while matching the distribution over lattice configurations of the energy level values, $E_\mathfrak{n}$, from which they were obtained.\footnote{The only exception to this are those levels lying `on-resonance', i.e.\ the four levels in the inset of Fig.~\ref{fig:rtilde}. The data correlations for these levels are handled slightly differently, reflecting the dominant sensitivity of $\tilde{r}_{\mathfrak{n}}(L)$ for these levels to the parameter describing the resonance coupling to $\pi\pi$. See Appendix~\ref{sec:LL_En_app} for details.}
This ensures that the data correlation for the $\tilde{r}_{\mathfrak{n}}(L)$ inherits (up to signs) the relatively mild data correlation of the corresponding $E_\mathfrak{n}$, in accordance with a linearised approximation to error propagation. Details of the implementation are presented in Appendix~\ref{sec:LL_En_app}.

\medskip 

Over the energy region of elastic scattering between the $\pi\pi$ threshold and the $K\overline{K}$ threshold, $\mathcal{F}(s)$ can be parameterized using a low-order polynomial in $s$,
\begin{equation}\label{eq:param_pol_s}
\mathcal{F}(s)/m_\pi^2  = \sum_{n=0}c_n \cdot \left( \frac{s-s_0}{s_0} \right)^n \, ,
\end{equation}
where we choose $\sqrt{s_0}= 0.135 \, a_t^{-1} = 2.85 \, m_\pi$, centering on the resonance peak, for convenience.

\begin{figure}
\includegraphics{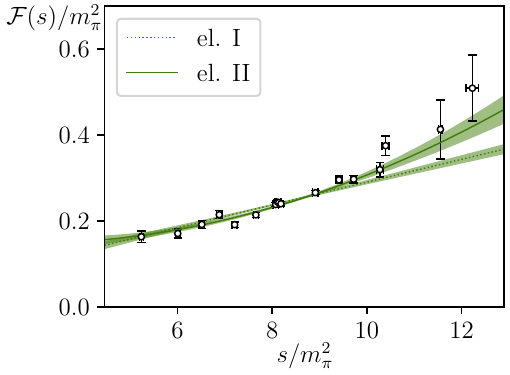}
\caption{\label{fig:sm_NovD_fit}Polynomial descriptions of $\mathcal{F}(s)$ -- linear in $s$ (el.\ I) and quadratic in $s$ (el.\ II).}
\end{figure}

As seen in Figure~\ref{fig:sm_NovD_fit}, fits linear or quadratic in $s$ can capture the observed energy dependence, with the quadratic fit having parameters,
\begin{center}
\renewcommand{\arraystretch}{1.2}	
    \begin{tabular}{rll}
    $c_0 =$                         & $0.2359\,(28)$  &
    \multirow{3}{*}{ $\begin{bmatrix*}[r] 1 & -0.3 & -0.6 \\
                                            & 1    &  0.7 \\
                                            &      & 1    \\ 
                                            \end{bmatrix*}$ } \\
    $c_{1}   = $                  & $0.265\,(26)$   & \\
    $c_{2} = $                  &  $ 0.20\,(7)$ & \\[1.3ex]
    \multicolumn{3}{l}{\quad\quad\quad\quad $\chi^2/N_{\text{dof}}=\frac{25.4}{17-3}=1.82$\,,}
    \end{tabular}
\end{center}\vspace{-0.9cm}
\begin{equation}\label{eq:el_II}\end{equation}
where parameter correlations are seen to not be excessive.
We note here that this fit has not imposed any additional constraints on the amplitude, such as the fact that the form factor must take value 1 at $s=0$ to reflect the charge of the pion. This is owing to the expectation that the simple \emph{reference elastic} amplitude parameterization will not be suitable for extrapolation far below the $\pi\pi$ threshold.
%

While the \emph{timelike} form factor, and its required finite-volume correction, has been our focus thus far, the \emph{spacelike} form factor can also be extracted on the same lattice.
A consistent description of both energy regions can be obtained by parameterizations with the appropriate analytic structure, e.g.\ the Omnès function, as we describe in the next section.

\section{Timelike \& spacelike pion form factor}\label{sec:TL_SL_ff}

The pion electromagnetic form factor in the \emph{spacelike} region can be extracted from calculations of three-point functions, $\langle 0 | \Omega_\pi(\Delta t) \, \mathcal{J}(t) \, \Omega^\dag_\pi(0) | 0 \rangle$, where the so-determined matrix-elements do not require significant finite-volume correction, as there is no two-hadron system that can go on-shell and sample the spatial lattice boundary. A description of our procedures for three-point function computation and analysis can be found in Ref.~\cite{Radhakrishnan:2022ubg} with some details presented in Appendix~\ref{sec:pi_3pt}. In this way, we determined the spacelike form factor over a range of $s <0$ (or equivalently $Q^2 = -s >0$ using the usual notation for virtualities) finding it to undergo monotonic decrease from value $1$ at $s=0$ (imposition of which serves to set the vector current renormalization factor as previously mentioned).

As discussed in Section~\ref{sec:production}, a \emph{dispersive} description of elastic $\pi\pi$ scattering can be constructed which is appropriate for analytic continuation from the timelike region into the spacelike region. Use of this `Omn\`es' form allows us to reliably connect our determinations of the spacelike and timelike form factors through a single amplitude. In brief, the relevant factor, $\Omega(s)$, in Eq.~\ref{eq:fOmnes_el} can be constructed using the elastic unitarity condition on the production amplitude, $f(s) - f^*(s) = 2i \, \mathcal{M}^*(s) \rho(s) f(s)$, and the Schwarz reflection principle to determine the discontinuity across the unitarity cut, ${f(s+i \epsilon) = e^{2i \delta(s+ i \epsilon)} \, f(s- i\epsilon)}$. A dispersive integral can be constructed which has the appropriate discontinuity,
\begin{equation*}
  \log \Omega(s+ i \epsilon) = \frac{s}{\pi} \int_{s_\mathrm{thr}}^\infty \! ds' \, \frac{\delta(s')}{s'\, (s' - s - i \epsilon)} \, ,
\end{equation*}
where a subtraction has been introduced that normalizes $\Omega(0) = 1$ while also suppressing the dependence on the elastic phase-shift at high energies.

Our \emph{reference elastic} scattering amplitude provides an elastic phase-shift from threshold up to $K\overline{K}$ threshold, and comparison with the \emph{reference coupled-channel} amplitude shows that the $\pi\pi$ phase-shift continues as given by the parameterization to somewhat higher energies. In practice we will use this form in the integral above up to $\sqrt{s_a} = 1.2 \cdot (2 m_K)$, and at energies higher than this a simple parameterization,
\begin{equation*}
  \delta( s > s_a) = \pi - \big(\pi - \delta_\mathrm{ref}(s) \big) \frac{2}{1+ \left(s/s_a\right)^{3/4} }\, ,
\end{equation*}
similar to the one proposed in Ref.~\cite{Colangelo:2018mtw}.\footnote{This form enforces an asymptotic return of the phase-shift to $\pi$, correcting for the threshold behavior built into the parameterization which ceases to be appropriate at high energies. In fact Ref.~\cite{Colangelo:2018mtw} used $\delta_\mathrm{ref}(s_a)$ in place of $\delta_\mathrm{ref}(s)$ in this expression but since the phase-shift of the \emph{elastic reference} parameterization has almost reached its asymptotic value at $\sqrt{s_a} = 1.2 \cdot (2 m_K)$, there is no practical difference between the two choices, and any small difference generated in $\Omega(s)$ can be absorbed into the smooth function $\mathcal{F}_\Omega(s)$.}
In practice the Omn\`es factor at low energies, where we require it, is not particularly sensitive to the details of this continuation, owing to the subtraction suppressing the contribution of high energies in the integral.

\begin{figure}
\subfloat[\label{fig:f_dat_sl}Spacelike region.]{\includegraphics{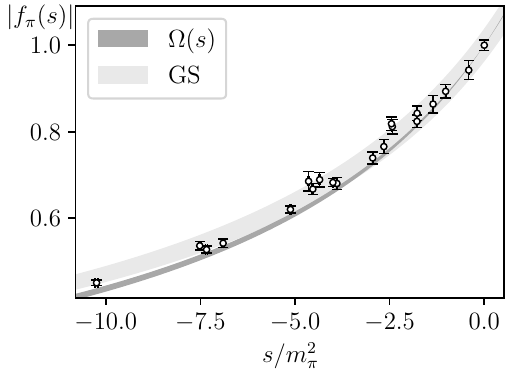}}\hfil
\subfloat[\label{fig:f_dat_sl+tl}Timelike region.]{\includegraphics{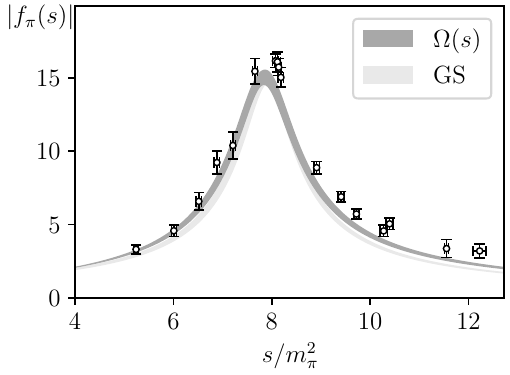}}
\caption{\label{fig:f_dat_Omnes} Pion form factor in the (a) spacelike and (b) timelike (as Fig~\ref{fig:f_abs_el}) regions. Superimposed are the Omnès function calculated from the \emph{reference elastic} amplitude, and the Gounaris-Sakurai (GS) form determined by describing the finite-volume spectra.
}
\end{figure}

The finite-volume corrected form factor data from Figure~\ref{fig:f_abs_el} is presented in Figure~\ref{fig:f_dat_Omnes}(b) with the Omn\`es factor, $\Omega(s)$, superimposed, where we clearly see that the bulk of the energy dependence is captured, with need only for a mild energy dependence in $\mathcal{F}_\Omega(s)$. Similarly Figure~\ref{fig:f_dat_Omnes}(a) shows our spacelike form factor data which also lies close to the (analytically continued) $\Omega(s)$ for $s<0$.

%
%
%

Also shown in these figures is a commonly-used parameterization known as the Gounaris-Sakurai form \cite{Gounaris:1968mw}, which effectively corresponds to an elastic $K$-matrix pole describing the $\rho$ resonance, along with a pole-subtracted Chew-Mandelstam phase-space.
The parameters of this amplitude were constrained by the lattice spectra in the elastic energy region.
The use of a dispersively improved phase-space removes the spurious singularity otherwise present at $s=0$ and thus makes an extrapolation into the space-like region somewhat plausible.

The function $\mathcal{F}_\Omega(s)$ obtained by dividing the Omn\`es factor out from the timelike and spacelike form factor data can be parameterized and a description of the lattice data obtained. Describing all data simultaneously requires spanning a large energy region, and it proves helpful to make use of a \emph{conformal mapping} of $s$ into a variable $z(s)$, defined as, 
\begin{equation} \label{eq:Zmap}
  z(s) = \frac{ \sqrt{s_c - s_0} - \sqrt{s_c - s} }{\sqrt{s_c - s_0} + \sqrt{s_c - s}} \, ,
\end{equation}
where this variable maps the entire complex plane of $s$, excluding the real $s$ axis above $s_c$, into a unit disk in $z$ centered so that $z(s_0) = 0$. Suitable choice of $s_c$ can reflect the fact that we expect there to be cuts due to inelastic channels at higher energies, and $s_0$ can be selected to conveniently distribute the data to be fitted around $z=0$. A polynomial in $z(s)$ will, as desired, have no singularities in the timelike elastic scattering region, or the spacelike region.

\begin{figure}[b]
\includegraphics{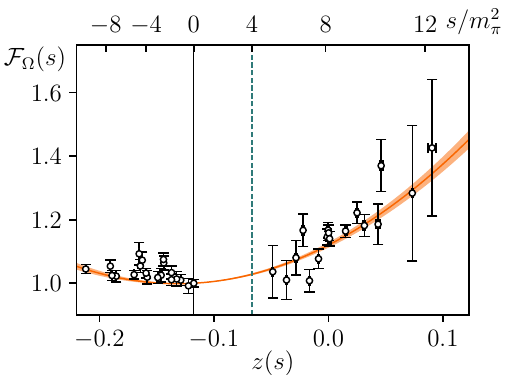}
\caption{\label{fig:sm_Omnes}
Ratio of the form factor data, $|f_\pi(s)|$ (as presented in Fig.~\ref{fig:f_abs_el}), to the Omnès function, $\Omega(s)$, as a function of the conformal map variable of Eq.~\ref{eq:Zmap}. The curve shows the polynomial fit of Eq.~\ref{eq:Omnes_params}. }
\end{figure}

\begin{figure}[b]
\subfloat[\label{fig:f_abs_sl+tl_all}Energy dependence across spacelike and timelike regions.]
{\includegraphics{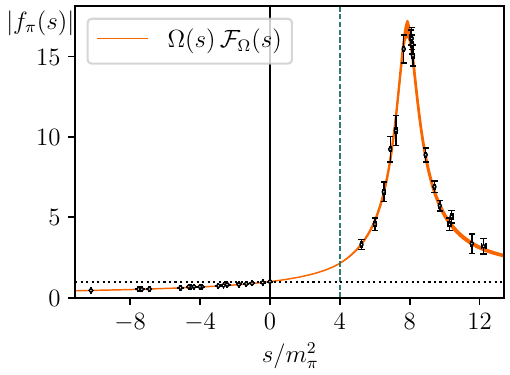}} \hfil
\subfloat[\label{fig:f_abs_sl+tl_sl}Zoom over the spacelike region.]{\includegraphics{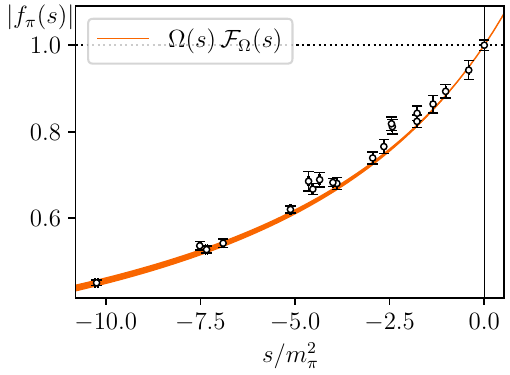}}
\caption{\label{fig:f_abs_sl+tl}Pion form factor across spacelike and timelike regions, and energy-dependent description by Omnès modulated by the quadratic $F_\Omega(s)$ of Eq.~\ref{eq:Omnes_params}.}
\end{figure}

Writing a parameterization of $\mathcal{F}_\Omega(s)$ as a low-order polynomial in $z(s)$,
\begin{equation}\label{eq:param_conf_m}
\mathcal{F}_{\Omega}(s) = Q_{\pi^+} + \sum_{n=1} d_n\cdot \big( z(s)^n-z(0)^n \big)\,,
\end{equation}
where the constraint from the fixed electric charge of the pion, $f_\pi(s\!=\!0)= \mathcal{F}_\Omega(s\!=\!0)= Q_{\pi^+} = 1$, is simply imposed, we can describe the lattice data across both spacelike and timelike regions, as shown in Figure~\ref{fig:sm_Omnes}. With $\sqrt{s_0} = 0.135\,  a_t^{-1}$ and $\sqrt{s_c} = 0.22 \, a_t^{-1} = 1.27\cdot (2m_K)$, the parameter values for a quadratic description are,
\begin{center}
\renewcommand{\arraystretch}{1.2}   
    \begin{tabular}{rll}
    $d_1 =$                         & $1.85\,(12)$  &
    \multirow{2}{*}{ $\begin{bmatrix*}[r] 1 & 0.8  \\
                                            & 1    \\ 
                                            \end{bmatrix*}$ } \\
    $d_{2}   = $                  & $7.0\,(5)$   & \\[1.3ex]
    \multicolumn{3}{l}{\quad\quad\quad\quad $\chi^2/N_{\text{dof}}=\frac{85.4}{37 - 2} = 2.44$\,.}
    \end{tabular}
\end{center}\vspace{-0.9cm}
\begin{equation}\label{eq:Omnes_params}\end{equation}
The corresponding pion form factor for this description of $\mathcal{F}_\Omega(s)$ is presented in Figure~\ref{fig:f_abs_sl+tl}. The somewhat large $\chi^2/N_{\text{dof}}$ is dominated by points in the spacelike region which have high statistical precision. Lattice discretization effects which are small in absolute terms, but of comparable size to the small statistical errors could explain this observation.

%
%
%

\begin{figure}
\includegraphics{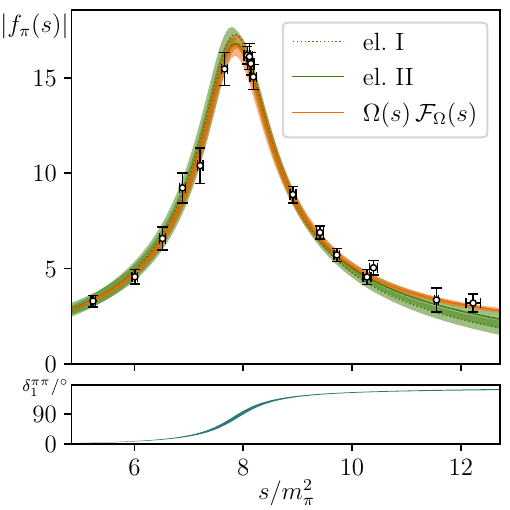}
\caption{
\label{fig:ff_el_fit}
Timelike pion form-factor described by \emph{reference elastic} amplitude via Eq.~\ref{eq:MF} (green) or via Omnès form (orange). These descriptions share a common phase given by the scattering amplitude phase-shift shown in the bottom panel.}
\end{figure}

Figure~\ref{fig:ff_el_fit} shows that over the timelike elastic scattering region, the Omn\`es approach and the previous approach using Eq.~\ref{eq:MF} yield compatible descriptions of the form factor.
It is clear that in the region of elastic scattering, the production amplitude can be determined with little systematic error. We now turn to the problem of determining production amplitude above the lowest inelastic threshold where we can produce real $K\overline{K}$ pairs.

\section{Timelike coupled-channel production amplitudes}

The additional constraint provided by the finite-volume matrix elements for discrete energy levels above $K\overline{K}$ threshold allows us to access the $\gamma \to K \overline{K}$ amplitude in addition to the $\gamma \to \pi\pi$ process at higher energies.
The extension of the finite-volume production formalism to multiple channels leads to a description of each finite-volume form-factor, $\mathcal{F}_\mathfrak{n}^{(L)}$, as a linear combination of the infinite-volume kaon and pion smooth production functions at $s=E^{\star2}_\mathfrak{n}$, as indicated in Eq.~\ref{eq:FVcorrec}.
Since there is no longer a one-to-one mapping from finite to infinite-volume, it is necessary to parameterize the energy dependence of the production functions, and globally describe multiple finite-volume form-factor values via a $\chi^2$ minimization, as proposed in Ref.~\cite{Briceno:2021xlc}.

\pagebreak

We implement the effect of the correlated uncertainty on the finite volume correction factors, $\tilde{r}_{\mathfrak{n},a}$, in terms of a `systematic' contribution to the data covariance in the fit $\chi^2$,
\begin{align*}
\chi^2 = \sum_{\mathfrak{n}, \mathfrak{m}} \Big[& \mathcal{F}^{(L)}_\mathfrak{n} - \big(\tilde{r}_{\mathfrak{n},\pi\pi} \mathcal{F}_{\pi\pi}(s) + \tilde{r}_{\mathfrak{n},K\overline{K}} \mathcal{F}_{K\overline{K}}(s) \big) \Big] \\
& \cdot \big( C^{\mathrm{stat.}} + C^{\mathrm{syst.}} \big)^{-1}_{\mathfrak{n}, \mathfrak{m}} \\
\cdot \Big[& \mathcal{F}^{(L)}_\mathfrak{m} - \big(\tilde{r}_{\mathfrak{m},\pi\pi} \mathcal{F}_{\pi\pi}(s) + \tilde{r}_{\mathfrak{m},K\overline{K}} \mathcal{F}_{K\overline{K}}(s) \big) \Big] \, ,
\end{align*}
which augments the `statistical' covariance of the $\mathcal{F}^{(L)}_\mathfrak{n}$. This `systematic' covariance is computed using the resampled $\tilde{r}_{\mathfrak{n},a}$ introduced in Section~\ref{sec:ff_tl_el} whose data covariance inherits that of the energy levels. For each energy level, only the dominant component, $a=\pi\pi$ or $K\overline{K}$ is used to compute the covariance. The detailed implementation is presented in Appendix~\ref{sec:LL_En_app}.  

\begin{figure*}
\includegraphics{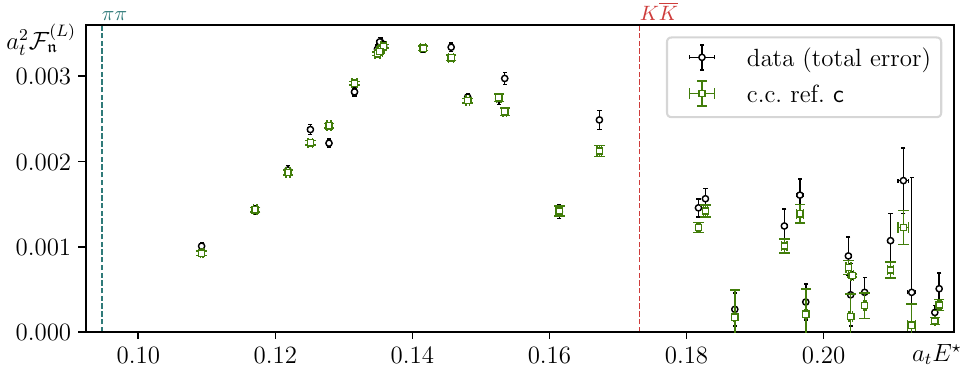}
\caption{\label{fig:FL_fits}
Finite volume form factor (open black circles) with errorbars modified to account for the fractional uncertainty on the finite-volume correction factors, coming from the diagonal elements of the `systematic' covariance, as described in the text. Global description using the \emph{reference coupled-channel} scattering amplitude, and order-1 polynomial forms for $\mathcal{F}_{\pi\pi}(s), \mathcal{F}_{K\overline{K}}(s)$ (green squares).
}
\end{figure*}

\vspace{2mm} 

To describe the infinite volume form factors we use the parameterization of Eq.~\ref{eq:fNovD}, with the \emph{reference coupled-channel} scattering amplitude determined in Sec.~\ref{sec:cc_scat}. The smooth functions $\mathcal{F}_a$ are parameterized with low order polynomials in $s$, analogous to Eq.~\ref{eq:param_pol_s},
\begin{equation}\label{eq:ccF}
\mathcal{F}_a(s)/m_\pi^2 = \sum_{n=0}^{N_a} \, h_{a,n}\cdot\qty(\tfrac{s-s_0}{s_0})^n\,.
\end{equation}
We emphasize here that the function $\mathcal{F}_{\pi\pi}(s)$ obtained in the coupled-channel case does not have to resemble the similarly named function in the \emph{elastic} case, presented in Sec.~\ref{sec:ff_tl_el}.
This can be seen by considering $f_\pi(s)$, which should be similar between the two cases, and which has a representation,
\begin{equation}\label{eq:comp_cc_fpi_MF}
f_\pi =  \frac{1}{k^{\star \, 2}_{\pi\pi}} \mathcal{M}_{\pi\pi,{\pi\pi}}\qty[\mathcal{F}_{\pi\pi} + 
\frac{k_{\pi\pi}^\star}{k_{K\overline{K}}^\star}\frac{\mathcal{M}_{\pi\pi,K\overline{K}}}{\mathcal{M}_{\pi\pi,{\pi\pi}}}\mathcal{F}_{K\overline{K}}]\,,
\end{equation}
in the coupled-channel case, where we have factored out the elastic $\pi\pi$ scattering in analogy to Eq.~\ref{eq:MF}. Below the $K\overline{K}$ threshold, the expression within square brackets is real and smooth (because the ratio of scattering amplitude components does not posses the $\pi\pi$ branch cut), and serves as the effective $\mathcal{F}$ in the prior elastic case.

Describing all the matrix element values previously presented in Figure~\ref{fig:FL_data} using finite-volume factors $\tilde{r}_{\mathfrak{n},a}(L)$ computed from the \emph{reference coupled-channel} amplitude and smooth functions $\mathcal{F}_{\pi\pi}, \mathcal{F}_{K\overline{K}}$ each described by polynomials of order 1, leads to the result summarized in Fig.~\ref{fig:FL_fits}. The description given by the green points is observed to be in good qualitative agreement with the constraining lattice data, with just a few isolated points having significant deviations which produce a somewhat large $\chi^2/N_{\text{dof}}$.

Examining Figure~\ref{fig:ff_el_fit} we see that the points which are most discrepant in the coupled-channel description were also somewhat discrepant in the elastic description, and that the larger effect in the current case may reflect the different approach for propagating the finite-volume factor uncertainty.

Table~\ref{tab:CC_funcforms} presents variations of both the polynomial orders used in $\mathcal{F}_{\pi\pi}(s)$, $\mathcal{F}_{K\overline{K}}(s)$, and the form used for the coupled-channel scattering amplitude, where we consider a second functional form of the $K$-matrix, given in Eq.~\ref{eq:cc_gamma_s}, which includes an additional linear term in $s$ with a set of $\gamma^{(1)}$ coefficients. The reduction in the $\chi^2$ values for the varied scattering amplitude can be entirely associated to the decreased precision of the finite-volume corrections factors, which is implemented as a larger systematic error on the data.

The final column of Tab.~\ref{tab:CC_funcforms} shows an alternative choice for the determination of the `systematic' covariance, where the magnitude of the diagonal elements are determined iteratively from the mean value of the model $\sum_a\tilde{r}_{\mathfrak{n},a} \, \mathcal{F}_a$. 
This prescription was inspired by the fitting procedure described in Ref.~\cite{McRae:2023zgu}, more details can be found in that reference and in Appendix~\ref{sec:LL_En_app}.
We found consistent results from both methods to determine the  `systematic' covariance, including the determination of the decay constant of the $\rho$-meson described in the following section, and shown in Fig.~\ref{fig:Vdcc}.

\begin{table}[]
\centering
\begin{tabular}{ccccccc}
\multirow{2}{*}{Model} & \multirow{2}{*}{$N_{\pi\pi}$} & \multirow{2}{*}{$N_{K\overline{K}}$ } &  \multirow{2}{*}{$N_{\text{dof}}$ } & \multicolumn{3}{c}{$\chi^2/N_{\text{dof}}$}\\
  & & & & reference & $\gamma^{(1)}$ & ref.\ iter. \\\hline
\textsf{a} & 1 & 0 & $32-3$ & 5.51 & 2.62 & 5.26 \\
\textsf{b} & 0 & 1 & $32-3$ & 4.39 & 2.41 & 4.90 \\
\textsf{c} & 1 & 1 & $32-4$ & 4.53 & 2.48 & 4.92 \\
\textsf{d} & 2 & 0 & $32-4$ & 4.87 & 2.55 & 5.28 \\
\end{tabular}
\caption{Global description of full set of finite-volume matrix elements. Variations in polynomial order in Eq.~\ref{eq:ccF} used with \emph{reference coupled-channel} scattering amplitude (fifth column), an alternative coupled channel amplitude with more parameter freedom, Eq.~\ref{eq:cc_gamma_s} (sixth column), and \emph{reference coupled-channel} amplitude with a modified iterative fitting strategy described in Appendix~\ref{sec:LL_En_app} (seventh column). }
\label{tab:CC_funcforms}
\end{table}

\begin{figure}
\includegraphics{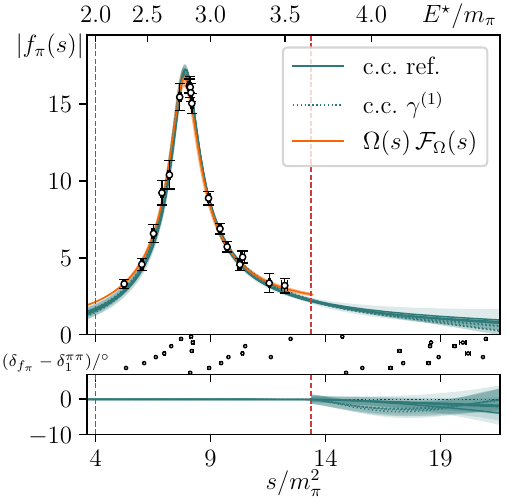}
\caption{\label{fig:fpi_cc}Timelike pion form factor using coupled-channel amplitude parameterizations showing variations detailed in Table~\ref{tab:CC_funcforms}(blue curves). Also shown the discrete elastic determinations of Section~\ref{sec:ff_tl_el} and the corresponding elastic Omnès description. The bottom panel shows the form factor phase difference with respect to the $\pi\pi$ phase-shift. The points in between panels show the energies of the levels used to constrain the form factor energy dependence.}
\end{figure}

The timelike pion form factor for these descriptions is shown in Figure~\ref{fig:fpi_cc}, where we see only modest spread over the different parameterizations, and rather close agreement with the result of Sec.~\ref{sec:TL_SL_ff} described in terms of the Omnès function over the elastic region.
The bottom panel of Fig.~\ref{fig:fpi_cc} shows the difference between the form factor phase and the $\pi\pi$ $P$-wave phase-shift, which above the $K\overline{K}$ threshold need not coincide\footnote{Watson's theorem equating the production phase to the scattering phase-shift applies rigorously only in the elastic region.}
However, due to the inelasticity in the scattering amplitude being close to 1, indicating little $\pi\pi, K\overline{K}$ channel coupling, we expect a small phase difference, and indeed we observe differences compatible with zero.

\begin{figure*}
\subfloat[\label{fig:f_aK.a}Pion timelike form factor.]{\includegraphics{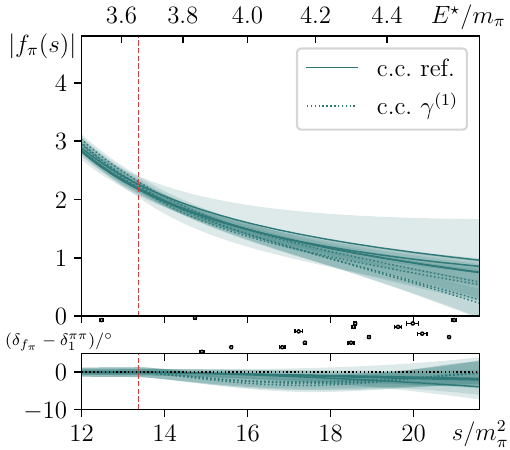}}
\hfil
\subfloat[\label{fig:f_aK.b}Kaon isovector timelike form factor.]{\includegraphics{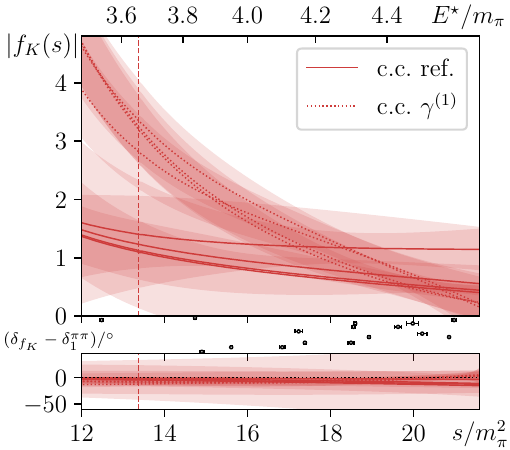}}
\caption{Coupled-channel parameterization variations of Table~\ref{tab:CC_funcforms} plotted over the inelastic energy region.}\label{fig:f_aK}
\end{figure*}

In Figure~\ref{fig:f_aK} we show the timelike pion and kaon (isovector) form-factors for energies above the $K\overline{K}$ threshold, where we observe a much greater sensitivity to the choice of scattering amplitude parameterization in the kaon case. In particular the kaon form-factor using the ``$\gamma^{(1)}$'' parameterization inherits a larger coupling to the $\rho$ from the scattering amplitude, and more of a resemblance to the tail of the resonance. Nevertheless, the kaon amplitudes broadly agree within the fairly large uncertainties. They differ most at and below $K\overline{K}$ threshold, but this is precisely where the finite-volume approach becomes unconstraining, as the contribution of the production amplitude to the finite volume matrix elements is exponentially suppressed below its threshold (see Appendix~A of Ref.~\cite{Briceno:2021xlc}).
In the current case, this can be understood in terms of the poorly constrained value of the coupling of the $K\overline{K}$ channel with the $\rho$-resonance. For a narrow resonance the value of $f_a$ at the resonance mass $m_R$ is approximated by,
\begin{equation}
f_a = -i \sqrt{16\pi}\frac{\hat{c}_a}{m_R\Gamma_R}\mathcal{F}^{(L)}_\mathfrak{n} + \mathcal{O}\left(\tfrac{\Gamma_R}{m_R}\right)\, ,
\end{equation}
and the value of $\hat{c}_{K\overline{K}}$ varies significantly between the \emph{reference coupled-channel} and the ``$\gamma^{(1)}$'' scattering parameterizations, with a greater value in the latter case. We report the resonance parameters of the ``$\gamma^{(1)}$'' scattering amplitude in Appendix~\ref{sec:scat_variation}.

\pagebreak
\section{
Pion and rho parameters from form factors}\label{sec:had_props}

The spacelike form factor is typically characterized by its slope at zero momentum transfer, what is commonly referred to as the `charge radius', $\ev{r_\pi^2}^{1/2}$, where\footnote{See Ref.~\cite{Jaffe:2020ebz}, and references therein, for a discussion about the proper interpretation of this quantity when it is comparable to the Compton wavelength of the system.}
\begin{equation}
\ev{r_\pi^2} \equiv 6 \eval{\frac{d }{d s}f_\pi(s)}_{s=0}\,.
\end{equation}
We determine this quantity from the parameterization in Eq.~\ref{eq:Omnes_params}, obtaining $\ev{r_\pi^2}^{1/2} = 0.614(7)$ fm.
This value sits in between the experimental pion charge radius, $\ev{r_\pi^2}^{1/2} = 0.659(4)$ fm, and kaon charge radius, $\ev{r_K^2}^{1/2} = 0.560(31)$ fm, reported by the PDG~\cite{Workman:2022ynf}, as we might expect given our lattice pion mass lying between the physical pion and kaon masses. This result is also consistent with previous analyses of timelike form factors determined at similar pion masses in lattice QCD~\cite{Andersen:2018mau,Erben:2019nmx}.

The timelike pion form-factor is dominated by the contribution of the $\rho$ resonance, and at the resonance pole, the $\pi\pi$ production amplitude factorizes into the coupling to the $\pi\pi$ system and to the electromagnetic current,
\begin{equation*}
\mathcal{H}^\mu_{\pi\pi,m} = \frac{\sqrt{16\pi}\, c_{\pi\pi}}{s_R-s} \, \big\langle \rho(\vec{P},m) \big| \mathcal{J}^\mu \big| 0 \big\rangle\,.
\end{equation*}
The current-vector matrix element can be expressed in terms of a Lorentz-invariant $\rho$-photon coupling, which we choose to parameterize in a dimensionless form,\footnote{Other normalizations have been used in the literature, see for example Ref.~\cite{Gasser:1982ap} that uses $f_\rho=\sqrt{2}\, m_R\, f_V$.}
\begin{equation} \label{eq:fV}
\big\langle \rho(\vec{P},m) \big| \mathcal{J}^\mu \big| 0 \big\rangle\ = \epsilon^{\mu*}(P,m)\, f_V \, m_R^2\,,
\end{equation}
where $m_R^2=\text{Re}(s_R)$.

One definition of the $\rho$ meson decay `partial width' into an electron-positron pair,
\begin{equation*}
\Gamma_{\rho\to e^+e^-} = \frac{4\pi \alpha^2}{3}m_R\abs{f_V}^2\,,
\end{equation*}
uses this coupling, and use of the PDG average for $\Gamma_{\rho\to e^+e^-}$ can thus provide an estimate for $|f_V|$.
A more consistent approach, which was taken in Ref.~\cite{Hoferichter:2017ftn}, would be to describe the $e^+ e^- \to \pi \pi$ cross-section energy dependence using the infinite-volume amplitude parameterizations presented earlier, and analytically continue them to the pole, yielding a complex-valued $f_V$.

Practical extraction of the $f_V$ coupling from our lattice-constrained amplitudes varies slightly depending upon the form of the amplitude construction. In the elastic case using $K$-matrix parameterizations, we may simply use the $\pi\pi$ coupling from the scattering amplitude and the singularity-free function, $\mathcal{F}(s)$, evaluated at the pole location~\cite{Hoferichter:2017ftn},
\begin{equation*}
f_V = -\sqrt{\frac{4}{3}} \frac{1}{m_R^2}\sqrt{16\pi}\,\hat{c}_{\pi\pi} \, \mathcal{F}(s_R)\,.
\end{equation*}
In the case of the Omnès parameterization of Eq.~\ref{eq:fOmnes_el} one needs to analytically continue the Omnès function $\Omega$ into the unphysical Riemann sheet.
This can be achieved by multiplying $\Omega$ by the $S$-matrix evaluated in the unphysical Riemann sheet where the scattering amplitude houses the resonance pole.

In the coupled-channel case, the extracted coupling value is independent of whether it is pulled from the pion or kaon form factor, in both cases being
\begin{equation}
\label{eq:fV_cc}
f_V =  -\sqrt{\frac{4}{3}} \frac{1}{m_R^2} \, \sqrt{16\pi}\, \Big(  \hat{c}_{\pi\pi}\mathcal{F}_{\pi\pi}(s_R)-\hat{c}_{KK}\mathcal{F}_{KK}(s_R) \Big)  \,.
\end{equation}

\begin{figure}
\includegraphics{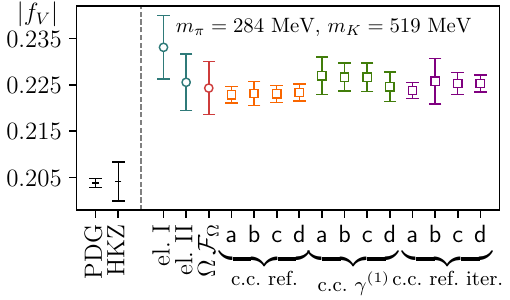}
\caption{\label{fig:Vdcc}Vector decay constant, $f_V$, for the $\rho$ meson. The first two values correspond to the experimental extraction reported by the PDG\cite{Workman:2022ynf} and the dispersive analysis of Hoferichter et.\ al.\ [HKZ] \cite{Hoferichter:2017ftn}. Subsequent values are the result of this analysis, for only the elastic region (circles) and also including the coupled channel region (squares).}
\end{figure}

In Figure~\ref{fig:Vdcc} we summarize our determinations of $f_V$ using extractions from elastic and coupled-channel parameterizations of various forms. The values are seen to be in good agreement, and as expected for a calculation at an unphysical quark mass, they differ somewhat from the experimental value following from the PDG's $\rho\to e^+e^-$ partial width.
The errorbars include the uncertainties on the resonance parameters, $c_a$ and $s_R$, extracted from the scattering amplitude, as well as those of the smooth functions $\mathcal{F}_a$, which are estimated from the variance of the $\mathcal{F}^{(L)}_\mathfrak{n}$ data and the finite-volume correction factors.
The orange and purple points show two different strategies for handling the error propagation from the finite-volume correction factors, as discussed in Appendix~\ref{sec:LL_En_app}.
Green points differ from the orange ones by the functional form of the scattering amplitude used to describe the spectra and to calculate the finite-volume factors, see Eq.~\ref{eq:cc_gamma_s} in Appendix~\ref{sec:scat_variation}.

The value coming from the Omnès parameterization serves as a conservative estimate at $m_\pi = 284$ MeV,
\begin{align*}
f_V&= 0.224(6) \, e^{-i\pi\cdot 0.0418(27)}\quad [-0.09]\,,\\{}
\Gamma_{\rho\to e^+e^-}&=8.9(5)\text{ keV}\,,
\end{align*}
where the coupling is observed to be close to being real valued, with only weak correlation between magnitude and phase.

A mild pion mass dependence of $f_V$ was observed in Ref.~\cite{Jansen:2009hr} where the $\rho$ was studied using only two light quark flavors and without explicit finite-volume correction (as it was either stable or very narrow at the pion masses considered), and in Refs.~\cite{Sun:2015enu,DeGrand:2019vbx} similarly for three flavors.
However, a rigorous study of the pion mass dependence of $f_V$ will require the implementation, as done in this work, of the finite-volume correction to the lattice matrix elements, and the analytic continuation to the resonance pole.
These effects become more significant as the pion mass decreases towards the physical point, because the decay phase-space to pions, hence the width of the resonance, increases, moving the resonance pole away from the real energy axis.

\section{Summary and outlook}

In this work we have presented a QCD calculation of the timelike form factor of the pion in a kinematic region where the final state interactions are sensitive to the rescattering of coupled $\pi\pi$ and $K\overline{K}$ systems. The formalism correcting for the finite spatial volume of the lattice necessarily gave us access to both the timelike pion and kaon form factors.

We also described \emph{simultaneously} the spacelike and the elastic timelike pion form factor with a dispersive representation, the Omnès function, satisfying the analyticity requirement of the amplitude in addition to the elastic unitarity requirement. This provides an improvement over the Gounaris-Sakurai form, which only satisfies the latter.

We found that considering a variety of functional forms for the energy dependence of the form factors, when determined from lattice matrix elements in the elastic timelike region, the spacelike plus elastic timelike region, and the timelike region including above the $K\overline{K}$ threshold, all gave consistent results for the decay constant of the $\rho$, the resonance which dominates the timelike pion form factor at low energies.
This observation provides a consistency check on the coupled-channel transition technology, motivating its application to more challenging cases of resonances decaying into multiple two-meson channels, e.g.\ the $\pi_1$ studied in Ref.~\cite{Woss:2020ayi}.

The current calculation was restricted to the \emph{isovector} component of the electromagnetic current, which is the only non-zero contribution in the case of the pion form factor. However, in the case of the electromagnetic \emph{kaon} timelike form factor, an isoscalar component is allowed, and phenomenologically this is known to be dominated at low energies by the isoscalar vector poles of the $\omega$ and $\phi$ resonances~\cite{Stamen:2022uqh}.
A calculation of the isoscalar component, in a lattice with pion masses $\lesssim 350$~MeV, would require a different technique to remove the power-law finite-volume corrections, given that the isoscalar $K\overline{K}$ channel couples to a system of \emph{three} pions. Progress on the development of this technology can be found in Refs.~\cite{Briceno:2017tce,Hansen:2021ofl}. %

The techniques employed here can also be applied to study electroweak processes involving multiple two-hadron final states.
For example, to calculate $CP$-violation in $D$-meson decays to $\pi\pi$ and $K\overline{K}$, which could be then compared to the recent experimental determination of Ref.~\cite{LHCb:2019hro}.


\begin{acknowledgments}
We thank our colleagues within the Hadron Spectrum Collaboration for their continued assistance. The authors acknowledge support from the U.S. Department of Energy contract DE-SC0018416 at William \& Mary, and contract DE-AC05-06OR23177, under which Jefferson Science Associates, LLC, manages and operates Jefferson Lab. 
This work contributes to the goals of the U.S. Department of Energy \emph{ExoHad} Topical Collaboration, Contract No. DE-SC0023598.

The software codes
{\tt Chroma}~\cite{Edwards:2004sx}, {\tt QUDA}~\cite{Clark:2009wm,Babich:2010mu}, {\tt QUDA-MG}~\cite{Clark:SC2016}, {\tt QPhiX}~\cite{ISC13Phi},
{\tt MG\_PROTO}~\cite{MGProtoDownload}, and {\tt QOPQDP}~\cite{Osborn:2010mb,Babich:2010qb} were used for the computation of the quark propagators.
The authors acknowledge support from the U.S. Department of Energy, Office of Science, Office of Advanced Scientific Computing Research and Office of Nuclear Physics, Scientific Discovery through Advanced Computing (SciDAC) program. 
Also acknowledged is support from the Exascale Computing Project (17-SC-20-SC), a collaborative effort of the U.S. Department of Energy Office of Science and the National Nuclear Security Administration.
This work was also performed on clusters at Jefferson Lab under the USQCD Collaboration and the LQCD ARRA Project.
This research was supported in part under an ALCC award, and used resources of the Oak Ridge Leadership Computing Facility at the Oak Ridge National Laboratory, which is supported by the Office of Science of the U.S. Department of Energy under Contract No. DE-AC05-00OR22725.
This research used resources of the National Energy Research Scientific Computing Center (NERSC), a DOE Office of Science User Facility supported by the Office of Science of the U.S. Department of Energy under Contract No. DE-AC02-05CH11231.
The authors acknowledge the Texas Advanced Computing Center (TACC) at The University of Texas at Austin for providing HPC resources.
Gauge configurations were generated using resources awarded from the U.S. Department of Energy INCITE program at the Oak Ridge Leadership Computing Facility, the NERSC, the NSF Teragrid at the TACC and the Pittsburgh Supercomputer Center, as well as at the Cambridge Service for Data Driven Discovery (CSD3) and Jefferson Lab.

\end{acknowledgments}

\bibliography{ps_TL_ff_refs.bib}

\appendix


\section{Spectrum sensitivity to use of \texorpdfstring{$\pi \omega$}{πω} operators}\label{app:spectrum}

The analysis of the $\pi\pi, K\overline{K}$ coupled channel system performed on this lattice in principle is only justified up to the next inelastic threshold, and the finite-volume approach we use is only rigorously correct up to the lowest $n > 2$ particle threshold.
However, in practice, if the $\pi\pi, K\overline{K}$ sector is decoupled from other channels as they open, the analysis can be carried out to higher energies.

The $\pi\omega$ and $\pi\pi\pi\pi$ thresholds are statistically consistent with each other, with the $\omega$ being effectively stable on this lattice. Experimentally the $\pi\pi\pi\pi$ channel is found to be weakly interacting up to 300 MeV above its threshold \cite{BaBar:2005dch}, and we expect the lack of $\pi\pi\pi\pi$-like operators to have a minimal impact in our analysis.
A similar situation occurs for the $\pi\pi\eta$ channel with a small cross-section at energies below the effective threshold for $\rho\eta$, so that the amplitude can be described by an isobar model \cite{BaBar:2007qju}, and the (negligible) impact of this channel in finite-volume was explored in a study similar to the present one in Ref.~\cite{Woss:2019hse}.

\begin{figure*}
\includegraphics[width=13cm]{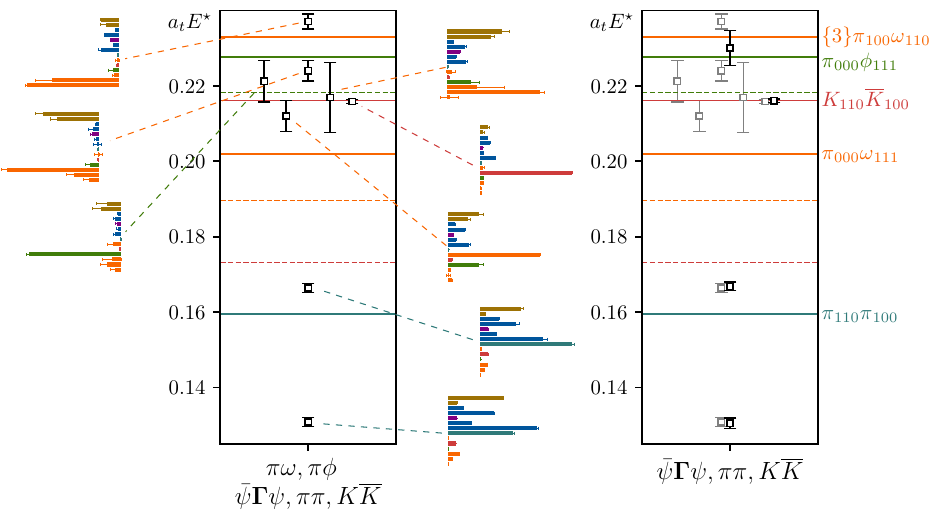}
\caption{Extracted finite-volume spectrum in the  $[111]\,E_2$ irrep from variational analysis of (left) correlation matrix including $\pi\omega$-like and $\pi\phi$-like operators, and (right) excluding such operators.
Dashed lines indicate meson-meson thresholds, and solid lines the non-interacting energy levels in this volume, with color coding as indicated below.
The histograms show the relative sizes of overlap factors for each state for each operator in the basis, color coded as: subduced single-meson with $J^P=1^-$ (dark blue), $J^P=1^+$ (brown), and $J^P=2^-$ (purple), $\pi\pi$ (teal), $K\overline{K}$ (red), $\pi\omega$ (orange) and $\pi\phi$ (green).
The spectrum from the left panel is reproduced on the right panel as the grey points to aid comparison.
}
\label{fig:omegapi_spec_op1}
\end{figure*}

Since in-flight irreps subduced from non-zero helicity contain contributions from $J^P=1^+$ as well as our desired $J^P=1^-$, the $\pi\omega$ can appear in $S$-wave, and as such we might anticipate it having a significant scattering amplitude\footnote{indeed we expect it to resonate as the $b_1$ in some energy region slightly higher than we have explored.}, and thus an impact on the finite-volume spectrum. 

We explicitly computed matrices of two-point correlations including operators resembling both $\pi \omega$ and $\pi \phi$ pairs, and carried out a variational energy-level extraction. An example of this is shown in Fig.~\ref{fig:omegapi_spec_op1}, showing  the extracted discrete spectrum in the $[111] E_2$ irrep including such operators (left), and excluding them (right).

As expected, the low-lying spectrum, well below the $\pi\omega$ threshold, is unaffected by the inclusion of the extra operators whose non-interacting energies lie much higher, and the overlaps, $Z_\mathfrak{n}^i=\mel{\mathfrak{n}}{\mathcal{O}_i^\dagger}{0}$, for these states with those operators are observed to be negligible.

Above the $\pi\omega$ threshold we can identify by the operator-state overlaps that some states are generated predominantly by the $\pi\omega$-like or $\pi \phi$-like operators.

Because the variational analysis approach is able to separate contributions to the matrix of correlation functions from multiple levels, even if they are almost degenerate, we can observe that the spectrum on the left contains a very precise energy level right on top of the $K_{110}\overline{K}_{100}$ non-interacting energy with dominant overlap onto a $K\overline{K}$-like operator, and which has  energy statistically consistent with another three levels overlapping dominantly with $\pi \omega$-like and $\pi \phi$-like operators. This $K\overline{K}$-like level remains essentially unchanged in an analysis that does not include the $\pi\omega$-like and $\pi \phi$-like operators, justifying their removal for the spectrum determination, and their further exclusion on the timelike form factor calculation. Comparable observations can be made in the other irreps considered in this paper.



\section{Parameterization variations of the scattering amplitude}\label{sec:scat_variation}

In order to determine the sensitivity of the extracted production amplitudes to the detailed description of scattering, we used a variety of parameterizations of the scattering amplitude to describe the finite-volume spectrum extracted from lattice QCD computed correlation functions.

A general feature of the amplitudes that are able to successfully describe the energy levels is a rapid phase-shift increase around $E^\star=0.135\, a_t^{-1}$, which is efficiently parameterized by including a pole term in the $K$-matrix (Eq.~\ref{eq:K_mat_el}). Our variations concern whether this pole is allowed to couple to the $K\overline{K}$ channel, what degree of polynomial is added to the pole, and whether a simple, or dispersively improved phase-space is used. Table~\ref{tab:el_params} summarizes three variations used for elastic analysis, and Table~\ref{tab:cc_variations} summarizes 23 variations in the coupled-channel case.

In the elastic case we found little variation in the scattering energy dependence from different functional forms, as illustrated in Fig.~\ref{fig:ps_el}.
In the coupled-channel analysis we find the parameterizations listed in Tab.~\ref{tab:cc_variations} in overall agreement, with slight variations seen on the energy dependence of the inelasticity, but statistically consistent values of the phase-shifts $\delta^{\pi\pi}_1$ and $\delta^{K\overline{K}}_1$.
The variations for the resonance pole position $s_R$ and coupling to the $\pi\pi$ channel $c_{\pi\pi}$ fall within the statistical uncertainty of the data.
However, as mentioned in the main text, the coupling to the $K\overline{K}$ channel is poorly determined in each parameterization, and some models differ from each other at the level of up to two standard deviations.

To illustrate the results with largest variation we select two of these coupled-channel amplitudes for propagation into the production amplitude analysis. The first is the reference amplitude of Eq.~\ref{eq:cc_ref_kmatrix}, the first entry of Tab.~\ref{tab:cc_variations}, while the second, which we label ``$\gamma^{(1)}$'', has functional form,
\begin{equation}\label{eq:cc_gamma_s}
\mathcal{K}_{ab}(s) = \frac{g_a \, g_b}{m^2 - s} +\gamma_{ab}^{(0)}+\gamma_{ab}^{(1)}s\,,
\end{equation}
corresponding to the tenth entry of Table~\ref{tab:cc_variations}.
The resonance pole of the ``$\gamma^{(1)}$'' parameterization is located at
\begin{equation}
a_t\sqrt{s_R} = 
0.1327(5) -\tfrac{i}{2}0.0096(4) \, ,
\end{equation}
consistent with Eq.~\ref{eq:sr_ref_cc}, while the channel couplings are,
\begin{align}
a_t\,c_{\pi\pi}        &= 0.0424(9)\, e^{-i \pi \cdot 0.055(4)}\,, \nonumber \\
a_t\,c_{K\overline{K}} &= 0.097(24) \, e^{i \pi \cdot 0.484(10)} \, ,\label{eq:gamma_s_coupling}
\end{align}
where we observe a noticeable variation on $c_{K\overline{K}}$ as compared to the reference amplitude case in Eq.~\ref{eq:coup_ref_cc}.
Even with this variation of the coupling, the value of $f_V$ determined with Eq.~\ref{eq:fV_cc} shows little dependence on the chosen scattering amplitude, see Fig.~\ref{fig:Vdcc}.

\begin{table}[h]
\centering
\begin{ruledtabular}
\begin{tabular}{lcc}
Parameterization                          & $N_{\text{pars}}$ & $\chi^2/N_{\text{dof}}$ \\ \hline
Relativistic Breit-Wigner                         & 2                 & 1.23                    \\
$\mathcal{K}=\frac{g^2}{m^2-s}$ (Gounaris-Sakurai) & 2                 & 1.29                    \\
$\mathcal{K}=\frac{g^2}{m^2-s} + \gamma$  (With $-i\rho$ phase space)         & 3                 & 1.23 \\   
$\mathcal{K}=\frac{g^2}{m^2-s} + \gamma$           & 3                 & 1.21                   
\end{tabular}
\end{ruledtabular}
\caption{\label{tab:el_params}$P$-wave elastic amplitude parameterization variations. Second and fourth entries use Chew-Mandelstam phase-space}
\end{table}

\begin{table}[]
\centering
\begin{ruledtabular}
\footnotesize
\begin{tabular}{llcc}
Parameterization & Restrictions  & $N_{\text{pars}}$ & $\chi^2/N_{\text{dof}}$ \\\hline
\multirow{5}{*}{$\mathcal{K}_{ab}=\frac{g_ag_b}{m^2-s} + \gamma_{ab}$}   & --                                       & 6    & 1.10 \\
                                                                         & $g_{K\overline{K}}=0$                    & 5    & 1.07 \\
                                                                         & $\gamma_{K\overline{K},K\overline{K}}=0$ & 5    & 1.07 \\
                                                                         & $\gamma_{\pi\pi,K\overline{K}}=0$        & 5    & 1.08 \\
                                                                         & $\gamma_{\pi\pi,\pi\pi}=0$               & 5    & 1.47 \\\hline
\multirow{4}{*}{$\mathcal{K}_{ab}=\frac{g_ag_b}{m^2-s} + \gamma^{(1)}_{ab}s$}  & --                                       & 6    & 1.05 \\
                                                                               & $g_{K\overline{K}}=0$                    & 5    & 1.03 \\
                                                                               & $\gamma^{(1)}_{K\overline{K},K\overline{K}}=0$ & 5    & 1.02 \\
                                                                               & $\gamma^{(1)}_{\pi\pi,K\overline{K}}=0$        & 5    & 1.02 \\\hline
\multirow{5}{*}{\begin{tabular}{l}$\mathcal{K}_{ab}=\frac{g_ag_b}{m^2-s} + \gamma_{ab}(s)$\\$\gamma_{ab}(s)=\gamma^{(0)}_{ab}+\gamma^{(1)}_{ab}s$\end{tabular}}   & --                      & 9    & 1.05 \\
                                                                         & $g_{K\overline{K}}=0$                          & 8    & 1.11 \\
                                                                         & $\gamma^{(1)}_{K\overline{K},K\overline{K}}=0$ & 8    & 1.07 \\
                                                                         & $\gamma^{(1)}_{\pi\pi,K\overline{K}}=0$        & 8    & 1.06 \\
                                                                         & $\gamma^{(1)}_{\pi\pi,\pi\pi}=0$               & 8    & 1.08 \\\hline
\multirow{4}{*}{\begin{tabular}{l}$\mathcal{K}_{ab}=\frac{g_a(s)g_b(s)}{m^2-s} + \gamma_{ab}$\\ $g_a(s)= g_a^{(0)}+g_a^{(1)}s$\end{tabular}}   & --       & 8    & 1.15 \\
                                                          &$g_{\pi\pi}^{(1)}=0$                    & 7   & 1.11 \\
                                                          &$\gamma_{\pi\pi,\pi\pi}=0$, $\gamma_{\pi\pi,K\overline{K}}=0$                     & 6   & 1.10 \\
                                                          &$\gamma_{\pi\pi,\pi\pi}=0$, $\gamma_{K\overline{K},K\overline{K}}=0$        & 6   & 1.11 \\
                                                          &$\gamma_{\pi\pi,K\overline{K}}=0$, $\gamma_{K\overline{K},K\overline{K}}=0$ & 6   & 1.10 \\\hline
\multirow{4}{*}{\begin{tabular}{l}$\mathcal{K}_{ab}=\frac{g_ag_b}{m^2-s} + \gamma_{ab}$\\ $I_{ab} = -i\,\delta_{ab}\,\rho_a$\end{tabular}}  & --  & 6 & 1.12 \\
                                                                               & $g_{K\overline{K}}=0$                    & 5    & 1.11 \\
                                                                               & $\gamma_{K\overline{K},K\overline{K}}=0$ & 5    & 1.12 \\
                                                                               & $\gamma_{\pi\pi,K\overline{K}}=0$        & 5    & 1.11 \\      
\end{tabular}
\end{ruledtabular}
\caption{$P$-wave coupled-channel amplitude parameterization variations.}
\label{tab:cc_variations}
\end{table}


\section{Data correlation between finite-volume correction factors} \label{sec:LL_En_app}

\begin{figure}[h]
\includegraphics[width=\linewidth]{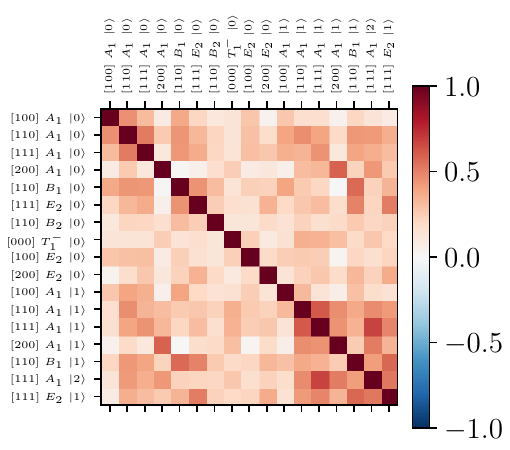}
\caption{\label{fig:correl_Ecm}
Data correlation matrix for the energy-level values, $E^\star_\mathfrak{n}$, with rows and columns ordered by increasing energy from left to right, and from top to bottom, respectively.}
\end{figure}

The $\tilde{r}_\mathfrak{n}(L)$ factors are calculated from scattering amplitudes described by a small number of parameters determined as ensembles over the set of lattice configurations.
This produces a high degree of correlation among them, which poses an implementation challenge to properly propagate their uncertainty into the determination of production amplitudes.
This high degree of correlation does not reflect the relatively modest amount of correlation between the energy levels in the finite-volume spectrum which constrain the scattering amplitude, as shown in Fig.~\ref{fig:correl_Ecm}.

%

In place of the data correlation for $\tilde{r}_{\mathfrak{n}}(L)$ computed using the amplitude parameterization, we adopt one which inherits the energy-level correlation, with the following motivation:
In a linearised approximation to error propagation, if the Jacobian were known, we could calculate the covariance of the $\tilde{r}_{\mathfrak{n}}(L)$ using
\begin{equation}\label{eq:Jac_LL_EE}
\sigma(\tilde{r}_{\mathfrak{n}},\tilde{r}_{\mathfrak{m}}) = \sum\nolimits_{\mathfrak{n}'\mathfrak{m}'}\pdv{\tilde{r}_{\mathfrak{n}}}{E_{\mathfrak{n}'}}\,
\sigma(E_{\mathfrak{n}'},E_{\mathfrak{m}'})\,
\pdv{\tilde{r}_{\mathfrak{m}}}{E_{\mathfrak{m}'}}\,.
\end{equation}
However, to a good approximation, we expect the finite-volume correlation factor computed with Eqns.~\ref{eq:slope}~and~\ref{eq:rtilde_expl} to be influenced predominantly by the \emph{local} behavior of $\mathcal{M}$ and $F$ around $E_\mathfrak{n}$.
This would imply that the Jacobian between $\tilde{r}_{\mathfrak{n}}(L)$ and the energies, $\{ E_{\mathfrak{n}'} \}$, is largely dominated by the diagonal elements, $\mathfrak{n}=\mathfrak{n}'$.

For instance, in an ideal scenario, where the lattice spectrum was determined for a large number of closely spaced volumes, $L$, and lattice irreps, $\Lambda$, we would have access to a high density of states per energy unit.
From this spectrum we could determine the scattering amplitude $\mathcal{M}(E)$ algebraically at a similarly high density, and
in that case, the calculation of $\tilde{r}_\mathfrak{n}(L)$ would not require an explicit parameterization of $\mathcal{M}$, and the uncertainty of each finite-volume correction factor would only be correlated to the uncertainty of the nearby energy levels.

\begin{figure}[h]
\includegraphics[width=\linewidth]{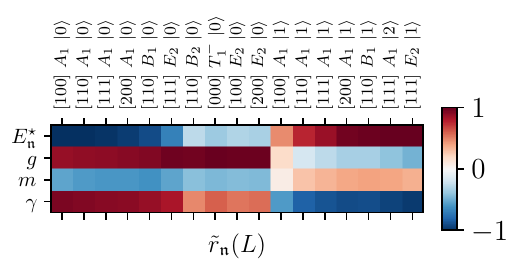}
\caption{\label{fig:diag_correl_Ecm_r}Data correlation between $\tilde{r}_\mathfrak{n}$ and the respective Lüscher energy solution $E^\star_\mathfrak{n}$, as well as the scattering parameters in the \emph{reference elastic} parameterization.}
\end{figure}

Based on this observation, we approximate the Jacobian of Eq.~\ref{eq:Jac_LL_EE} by a diagonal matrix $\pdv{\tilde{r}_{\mathfrak{n}}}{E_{\mathfrak{n}'}}\propto \delta_{\mathfrak{n},\mathfrak{n}'}$.
In practice, we implement this procedure by shifting and reweighing the jackknife ensemble of each energy level to match the mean and error of the respective finite-volume correction factor according to the values computed using the parameterization of $\mathcal{M}$ and Eqns.~\ref{eq:slope},~\ref{eq:rtilde_expl}.
With this prescription it is not necessary to compute the magnitude of the Jacobian elements, however we still require knowledge of their sign.
These are extracted from the correlation between the `model' energy, i.e.\ the solution of Eq.~\ref{eq:Luscherqc}, and $\tilde{r}_{\mathfrak{n}}(L)$, both extracted from a parameterization of $\mathcal{M}$~\footnote{Support for this prescription is found empirically in the fact that the absolute value of this correlation value is consistent with 1 for most energy levels.}, these are shown on the top row of Figure~\ref{fig:diag_correl_Ecm_r}.

An exception to this behavior are those levels with energies very close to the resonance mass, i.e.\ the four levels in the inset of Fig.~\ref{fig:rtilde}. As mentioned before, the value of $\tilde{r}_\mathfrak{n}(L)$ for these levels is equal to the resonance coupling, modulo rather small finite-width corrections, and we observe that for these levels $\tilde{r}_\mathfrak{n}(L)$ is mostly uncorrelated with the respective energy value. This is illustrated in Figure~\ref{fig:diag_correl_Ecm_r}, where we see that their strongest correlation (for the \emph{reference elastic} parameterization) is with the $g$ parameter. From the spectra of Fig.~\ref{fig:full_spec} we note that these levels are the farthest from any non-interacting energy level, and as such they are the least sensitive to finite-volume effects.

To account for this observation, for these levels we adopt a modified procedure, where instead of only using the energy level data, we add a linear combination of the fluctuations on the scattering parameters, with the parameter $g$ having the highest contribution.
For $\tilde{r}_\mathfrak{n}(L)$ for `on-resonance' levels, we form a linear combination of the ensembles of the scattering parameters and the corresponding energy level,
\begin{equation}
\vec{\tilde{r}}_\mathfrak{n}(L) = x_E \,\vec{E}_{\mathfrak{n}} + x_g \,\vec{g} +x_m \,\vec{m} + x_\gamma \,\vec{\gamma} =\sum\nolimits_ix_i\, \vec{p}_i.
\end{equation}
where we use the vector notation $\vec{p}_i$ to emphasize that we are manipulating the quantity $a$ in the ensemble configuration space. 
The quantities $\vec{a}$ have been previously shifted and reweighed, so that they have zero mean and unit covariance, i.e.\ $\ev{\vec{p}_i}=0$ and $\ev{\vec{p}_i\,^2}=1$.

To obtain the coefficients $x_i$ we demand that the correlation between the generated $\tilde{r}_\mathfrak{n}$ ensemble and each of the ${p}_i$ quantities reproduces the correlation $\text{corr}(p_i,\tilde{r}_\mathfrak{n}(L))$ shown in Fig.~\ref{fig:diag_correl_Ecm_r}.
Mathematically, this translates into the system of equations
\begin{align}
\sum\nolimits_j\ev{\vec{p}_i \,  \vec{p}_j} x_j &= \text{corr}(p_i,\tilde{r}_\mathfrak{n}(L))\,,\label{eq:x_params}
\end{align}
where the matrix $\ev{\vec{p}_i \,  \vec{p}_j}$ is calculated using the corresponding lattice energy level and the scattering parameters.
Equation~\ref{eq:x_params} represents a linear system of equations from which the coefficients $x_i$ can be easily obtained.
We find that the value of $\ev{\vec{\tilde{r}}_\mathfrak{n}(L)^2}$ extracted in this way is equal to unity, providing a self-consistency check of this procedure.
Finally we reweight and shift $\vec{\tilde{r}}_\mathfrak{n}(L)$ to obtain a jackknife ensemble with the mean and error of $\tilde{r}_\mathfrak{n}(L)$.

\vspace{2mm} 

It is not immediately obvious how to extend the procedure we just described from the elastic analysis to the coupled channel case, given that multiple components $\tilde{r}_{\mathfrak{n},a}(L)$ are associated with each level.
Furthermore, we need the correlation between the components of $\tilde{r}_{\mathfrak{n}}$ to reproduce the correlation between different elements of the scattering amplitude.
For instance, to stay consistent with unitarity the following relation (see Appendix~A1 of Ref.~\cite{Briceno:2021xlc}) needs to be satisfied in the elastic energy region:
\begin{equation}
\frac{\tilde{r}_{\mathfrak{n}, K\overline{K}}}{\tilde{r}_{\mathfrak{n}, \pi\pi}}
=
\frac{k_{\pi\pi}^{\star} \, \mathcal{M}_{\pi\pi,K\overline{K}}}
{k_{K\overline{K}}^{\star}\, \mathcal{M}_{\pi\pi,\pi\pi}}\,.
\end{equation}

We instead choose to estimate the impact of the uncertainty of the finite-volume correction by means of a ``systematic'' covariance, $C^{\mathrm{syst.}}$, applied to the finite-volume matrix elements $\mathcal{F}_\mathfrak{n}^{(L)}$.
We add this systematic covariance matrix to the statistical covariance matrix of $\mathcal{F}_\mathfrak{n}^{(L)}$ to obtain a total covariance, which in turn is used for the $\chi^2$ function used to determine the smooth functions $\mathcal{F}_a(s)$.

To calculate the magnitude of the diagonal elements of this covariance we first note that for levels below the $K\overline{K}$ threshold the component $\tilde{r}_{\mathfrak{n},K\overline{K}}$ cannot be precisely determined (by the nature of the finite-volume formalism).
Therefore, we assign the relative uncertainty of the ``relevant'' $\tilde{r}_{\mathfrak{n},\pi\pi}$ as the systematic relative uncertainty of the corresponding $\mathcal{F}_\mathfrak{n}^{(L)}$.
We note that following this procedure for the elastic analysis we obtain results consistent with the alternative procedure of replacing the correlation of the finite-volume correction factors by that of the energies and propagating the error over jackknife.

Above the $K\overline{K}$ threshold we have a weakly interacting system, where each state in the spectrum is close to a non-interacting energy.
This allows us to unambiguously assign a dominant component $\tilde{r}_{\mathfrak{n},a}(L)$ to each level.
Furthermore, following the expectations outlined about the correlation between the finite-volume correction factor and the energy value, we find that out of the two, this component is always the most correlated with the energy solution of the quantization condition.
Once we picked the dominant component $a$, we again assign the relative uncertainty of $\tilde{r}_{a,\mathfrak{n}}$ as the ``systematic'' relative uncertainty of the corresponding $\mathcal{F}_\mathfrak{n}^{(L)}$.

Finally, to obtain off-diagonal elements of $C^{\mathrm{syst.}}$, showing the correlation of the finite-volume correction across the spectrum, we mimic the prescription used in the elastic analysis.
This means that we will once again use the correlation matrix of the energy levels, or the appropriately constructed linear combinations for the levels close to the resonance.
This correlation matrix is multiplied by the magnitude of the diagonal elements determined above to obtain $C^{\mathrm{syst.}}$.

\vspace{2mm} 

We also explored an alternative procedure to determine the diagonal elements of $C^{\mathrm{syst.}}$ to corroborate the consistency of our prescription.
This second option was inspired by the iterative fit method described in Ref.~\cite{McRae:2023zgu}.
In this case we use the value of the ``model'', $\sum_a\tilde{r}_{a,\mathfrak{n}}(L)\mathcal{F}_a(E_\mathfrak{n}^{\star2})$, to multiply the relative uncertainty of the finite-volume correction factor and obtain an absolute systematic error for each diagonal element of $C^{\mathrm{syst.}}$.
In the first iteration we pick the solution from our previous prescription to determine $C^{\mathrm{syst.}}$ and minimize the $\chi^2$ with this systematic covariance.
This yields new values for $\mathcal{F}_a(s)$, which are then used to recompute $C^{\mathrm{syst.}}$ and repeat the minimization.
We find that this process converges after a few iterations.
The results of this iterative procedure show consistency with the previous choice of using $\mathcal{F}_\mathfrak{n}^{(L)}$ to determine the magnitude of the systematic errors.
This can be seen in the very similar values of the $\chi^2$ for each option, shown in Table~\ref{tab:CC_funcforms}, and the determination of the decay constant, $f_V$, shown in Figure~\ref{fig:Vdcc}.



\section{Subduction of kinematic factor \texorpdfstring{$K(\Lambda)$}{K(Λ)}}\label{sec:kin_fac_subd}

To obtain the kinematic factor $K(\Lambda)$ we need to project the components of the Lorentz vector $K^\mu_m$ of Eq.~\ref{eq:pw_kin} into a basis of definite helicity, to then be able to apply the subduction coefficients $\mathcal{S}^\lambda_\Lambda$ of Table~II of Ref.~\cite{Thomas:2011rh}. To illustrate this procedure, we will focus on the polarization vector $\epsilon^\mu(P,m)$, and include all other scalar factors at the end.

We begin by calculating the components of the polarization vector in the frame of momentum $P^\mu$, with a final hadron pair in a state of definite helicity $\lambda$.
This is achieved by boosting the rest-frame polarization vector, with components $(0,\epsilon^i_\lambda)^\mu$, to the frame of $P^\mu$ in a two-step process: first a boost is applied in the $z$-direction with velocity $\beta=\abs{\mathbf{P}}/P^0$, followed by a rotation $R_{\hat{\mathbf{P}}}$ from the $z$-axis to the direction of the spatial part of $P^\mu$.
Applying these transformations to the rest-frame polarization vector we obtain
\begin{equation*}
\epsilon^\mu(P,\lambda) =\begin{pmatrix}
\gamma\beta \, \delta_{\lambda ,0} \\
\epsilon^{i}_\lambda(\hat{\mathbf{P}}) \big(\gamma \, \delta_{\lambda,0} + \delta_{\lambda,\pm} \big)
\end{pmatrix}\,,
\end{equation*}
where $\epsilon^{i}_\lambda(\hat{\mathbf{P}})\equiv[R_{\hat{\mathbf{P}}}]^i{}_j \epsilon^{j}_\lambda$, and $\gamma$ is the relativistic factor associated to $\beta$.

Once we know the cartesian components of the polarization vector, we follow Ref.~\cite{Thomas:2011rh} to project the different $\mu$ components of $K^\mu_\lambda$, according to their properties under spatial rotations, into a helicity basis with spin-components $\hat{\lambda}$.
This process converts $K^\mu_\lambda$ into a diagonal matrix in the $\{\hat{\lambda},\lambda\}$ space, with the temporal component only spanning elements with zero helicity, while the spatial components span helicities $0$~and~$\pm1$,
\begin{align*}
K^{0}_\lambda &\to K^{[0]}_{\hat{\lambda}\lambda}= \sqrt{\tfrac{4}{3}}  \, \gamma\beta\,   \delta_{\hat{\lambda},0}\,\delta_{\lambda,0}\,\,, \\
K^{i}_\lambda &\to K^{[i]}_{\hat{\lambda}\lambda}= \sqrt{\tfrac{4}{3}}\, \qty( \gamma \delta_{\hat{\lambda},0}\,\delta_{\lambda,0} +   \delta_{\hat{\lambda},\pm}\, \delta_{\lambda,\pm} )\,.
\end{align*}
Finally, the kinematic factor $K(\Lambda)$ is calculated by applying the appropriate subduction coefficients 
\begin{equation*}
\sum\nolimits_{\hat{\lambda},\lambda} \mathcal{S}_{\Lambda}^{\hat{\lambda}}\,K^{[\mu]}_{\hat{\lambda}\lambda}\,\mathcal{S}_{\Lambda'}^{\lambda} = \delta_{\Lambda\Lambda'}K(\Lambda)\,,
\end{equation*}
the result of which listed in Table~\ref{tab:kin_fac}.
This last step is actually trivial because the subduction matrices are unitary, and the matrix $K^{[\mu]}_{\hat{\lambda}\lambda}$ is proportional to the identity in the subspace where the subduction matrices are non-zero.

\begin{table}
\centering
\begin{tabular}{c|cccc}
$\Lambda(\mathcal{J}^\mu)$ & $T_1^-(\mathcal{J}^i)$             & $A_1(\mathcal{J}^0)$                          & $A_1(\mathcal{J}^i)$                     & $E_2,B_1,B_2(\mathcal{J}^i)$ \\ \hline
$K(\Lambda)$ & $\sqrt{\tfrac{4}{3}}$ & $\sqrt{\tfrac{4}{3}}\gamma\beta$ & $\sqrt{\tfrac{4}{3}}\gamma$ & $\sqrt{\tfrac{4}{3}}$         \\
\end{tabular}
\caption{\label{tab:kin_fac}Kinematic factor of the subduced finite volume matrix elements into irrep $\Lambda$ from the temporal, $\mathcal{J}^0$, or spatial, $\mathcal{J}^i$, components of a vector current.}
\end{table}



\section{Timeslice fit algorithm of the finite-volume matrix elements}\label{sec:fit_algo}

Our approach to extracting a single value of the finite-volume matrix element from timeslice correlation functions, as presented in Fig.~\ref{fig:mel_td_fit}, follows the `model averaging' prescription suggested by Jay and Neil~\cite{Jay:2020jkz}.

Insisting that a minimal number of timeslices $N_t$ must be used in any fit, we perform fits to a constant $C$ with varying values $t_{\text{min}}$ but a fixed value of $t_{\text{max}}$.
For each such fit we compute a version of the Akaike Information Criterion (AIC), by combining the correlated $\chi^2$ and the number of degrees of freedom, $N_{\text{dof}} = N_\mathrm{data} - N_\mathrm{params}$, according to ${w = \exp[-(\chi^2/2 - N_{\text{dof}})]}$.

We label $t^\star$ the value of $t_{\text{min}}$ having the maximum value of $w$, which according to the criterion identifies this range as the dataset for a given $t_{\text{max}}$ best described by a constant.
Subsequently fits are attempted for all $t_{\text{min}} < t^\star$ using a constant plus an exponential.

We repeat this procedure for all values of $t_{\text{max}}$ allowed by $N_t$, yielding a large number of fits to a constant or a constant plus an exponential for many timeslice fit windows.
Each fit, $\alpha$, yields a value and a statistical error for the constant, $C_\alpha \pm \sigma_{C,\alpha}$, which can be ranked according to their respective value of AIC weight $w_\alpha$, and a \emph{model average} computed from a weighted average of a representative set of the fits with the largest weight,
\begin{align}
C &=  \frac{\sum_\alpha w_\alpha  C_\alpha }{\sum_\beta w_\beta }\,, \label{eq:mod_avg}\\
\sigma_{C}^2 &= \frac{\sum_\alpha w_\alpha  \sigma_{C,\alpha }^2 }{\sum_\beta w_\beta }
+ \frac{\sum_{\alpha \beta}w_\alpha w_\beta (C_\alpha -C_\beta)^2}{2\qty(\sum_\beta w_\beta)^2}\,. \label{eq:mod_avg_err}
\end{align}

To make this process compatible with the jackknife resampling technique used for error propagation, we employ the following prescription, based on the method presented in Ref.~\cite{Segner:2023a0}.
For each fit we can obtain an ensemble $\{C_{\alpha,i}\}$, where the index $i$ indicates an entry in the jackknife ensemble of model $\alpha$.
Then we can define the model averaged ensemble,
\begin{equation*}
C_i  = \frac{\sum_\alpha w_\alpha \, C_{\alpha,i}}{\sum_\beta w_\beta }\,.
\end{equation*}
The average of this ensemble is equal to the value given in Eq.~\ref{eq:mod_avg}, but the variance is bounded from above by the value of Eq.~\ref{eq:mod_avg_err} (typically we find it to be a few percent lower).
We can fix this mismatch by adding Gaussian noise to $C_i$, as long as the random variable $\eta_i$ has zero mean and a variance such that the variance of
\begin{equation}\label{eq:C_p_eta}
\overline{C}_i = C_i + \eta_i\,
\end{equation}
matches Eq.~\ref{eq:mod_avg_err}.
The Gaussian noise variables, $\eta_i$, are drawn from a multidimensional uncorrelated Gaussian distribution, such that this prescription does not impact the covariance among different matrix elements.
Reference~\cite{Jay:2020jkz} does not address the model average of the covariance among different variables, but we assume that it is best to not modify it.
We employ the ensembles of Eq.~\ref{eq:C_p_eta} to calculate the form-factors presented in this work.



\section{Late time pollution to matrix elements}\label{sec:late_t_poll}

The use of optimized operators at the source in correlation functions having the vector current at the sink allows us access to finite-volume matrix elements for states above the ground state owing to their overlap with other states in the spectrum being highly suppressed, $\mel{\mathfrak{m}}{\Omega_{\mathfrak{n}}^\dagger(0)}{0} \ll 1$, for $\mathfrak{m}\neq\mathfrak{n}$. 
This is a powerful technique~\cite{Shultz:2015pfa}, but we do encounter cases where a hierarchy of current matrix elements for different states can compensate for this suppression, i.e.\ where ${\mathcal{F}_\mathfrak{m}^{(L)}\cdot \mel{\mathfrak{m}}{\Omega_{\mathfrak{n}}^\dagger(0)}{0} \sim \mathcal{F}_\mathfrak{n}^{(L)}}$.

For those cases the ratio of correlation functions of Eq.~\ref{eq:Rdef} has non-negligible additive contributions from states other than $\mathfrak{n}$ of the form,
\begin{equation}\label{eq:x-talk_term}
\varepsilon_{\mathfrak{m},\mathfrak{n}}(t) = \varepsilon_{\mathfrak{m},\mathfrak{n}} \,  e^{(E_\mathfrak{n}-E_\mathfrak{m})t}\,,
\end{equation}
such that states lighter than $\mathfrak{n}$ will cause a rising time-dependence, visible at late times, that is not accounted for in our default timeslice fitting form.

Use of a GEVP solution at an appropriately large value of $t_0$ to form the optimized operators places some constraints on the scale of these late-time pollutions. The optimized correlation function matrix, $\mel{0}{\Omega_\mathfrak{m}(t)\Omega_{\mathfrak{n}}^\dagger(0)}{0}$, is diagonal at $t=t_0$, and close to diagonal for timeslices close to $t_0$.\footnote{In practice this matrix is not \emph{exactly} diagonal because the optimized operators are constructed with the ensemble averaged vectors $v_\mathfrak{n}$.}
The two-point current correlation functions, $\mel{0}{\mathcal{J}(t)\Omega^\dagger_\mathfrak{n}(0)}{0}$, are calculated on the same ensemble, and with the same time sources as the matrix of correlators used in the GEVP, leading to a significant correlation between the contributions $\varepsilon_{\mathfrak{m},\mathfrak{n}}(t)$ and the off-diagonal elements of the matrix of optimized correlators, suggesting that $\varepsilon_{\mathfrak{m},\mathfrak{n}}(t)$ will be also suppressed for timeslices around $t_0$~\footnote{Contributions from terms like Eq.~\ref{eq:x-talk_term} to $R_{\mathfrak{n}}(t)$ are not relevant for $t<t_0$, where pollution from states \emph{higher} in energy come to dominate correlation functions.}.

In the analysis reported in this paper, the only cases found where an excited state energy level has a current matrix element that is significantly smaller in magnitude than the current matrix element for a lighter state, and hence where this `late-time' pollution needs to be considered, are those energy levels lying close to non-interacting $K\overline{K}$ energies and having dominant overlap onto $K\overline{K}$-like operators.
For these levels, based upon the logic above, we restrict the time-windows for constant and constant-plus-exponential fit forms to more modest values of $t_{\text{max}} \sim 1.5\, t_0$, thus excluding late-times where the pollution from lighter states begins to become significant.
In practice, we only need to impose this fitting-window restriction explicitly for three states, the second excited energy level of each of the irreps $[100]\,A_1$, $[110]\,A_1$, and~$[111]\,E_2$. Four other states lying close to non-interacting $K\overline{K}$ energies in other irreps show a modest late-time enhancement of the form suggested by Eq.~\ref{eq:x-talk_term}, but the AIC-value quantifying the quality of fits favored those with lower values of $t_{\text{max}}$ anyway.

\bigskip

We present below an example case supporting the arguments presented above, in which we will reconstruct the time-dependence of correlation functions in terms of contributions of the form of Eq.~\ref{eq:x-talk_term}.
For convenience of presentation we will suppress kinematic factors not relevant to the illustration, by introducing unit-normalized optimized operators $\mel{\mathfrak{n}}{\Omega^{\prime\dag}_\mathfrak{m}(0)}{0} = \delta_{\mathfrak{n},\mathfrak{m}}$, the ratio,
\begin{equation*}
R^\prime_{\mathfrak{n}}(t) \equiv \frac{\langle 0 | \mathcal{J}(t)^{}\, \Omega^{\prime\dag}_\mathfrak{n}(0) |0\rangle}
{\langle 0 | \Omega^{\prime}_\mathfrak{n}(t) \, \Omega^{\prime\dag}_\mathfrak{n}(0) |0\rangle}\,,
\end{equation*}
and constants $\mathfrak{f}^{(L)}_\mathfrak{n}$ describing the leading constant time-dependence of ${R}^\prime_{\mathfrak{n}}(t)$.

\begin{figure*}
\subfloat[\label{fig:mel_KK_lvl0}Level 0.]{\includegraphics[width=.4\linewidth]{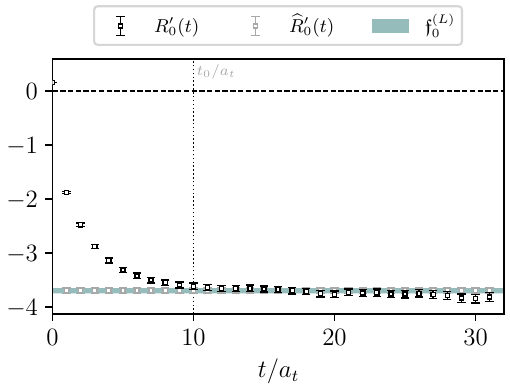}}
\hfil
\subfloat[\label{fig:mel_KK_lvl1}Level 1.]{\includegraphics[width=.4\linewidth]{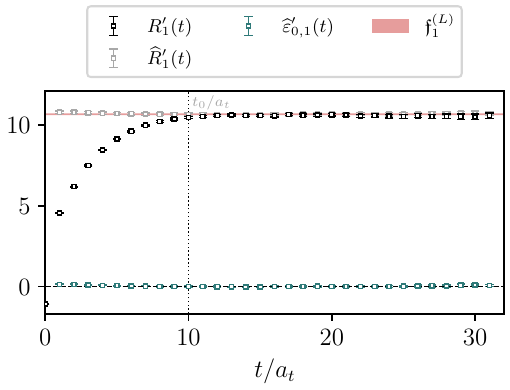}}

\subfloat[\label{fig:mel_KK_lvl2}Level 2.]{\includegraphics[width=.4\linewidth]{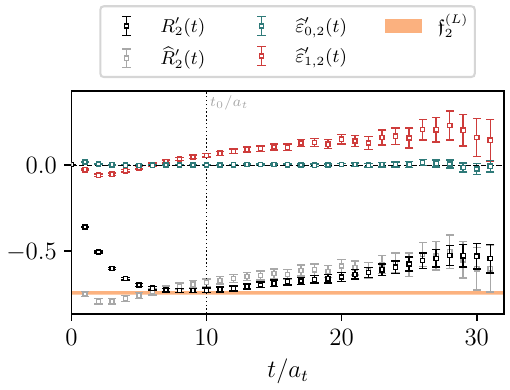}}
\hfil
\subfloat[\label{fig:mel_KK_lvl3}Level 3.]{\includegraphics[width=.4\linewidth]{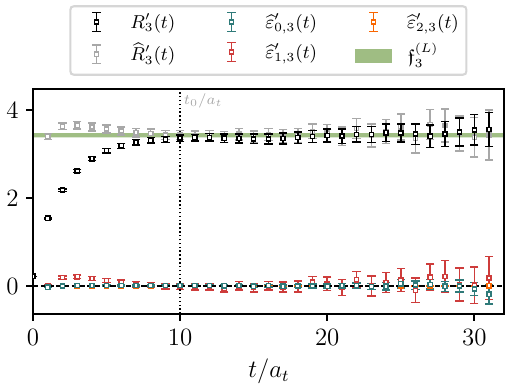}}
\caption{\label{fig:mel_KK_recon}Ratio of two-point correlators $R^\prime_\mathfrak{n}(t)$ (unimproved current) for discrete states in the $[100]\, A_1$ irrep compared with the extracted constants $\mathfrak{f}^{(L)}_\mathfrak{n}$, the expected contributions from lower energy levels, $\widehat{\varepsilon}^\prime_{\mathfrak{m},\mathfrak{n}}(t)$, and the sum of all contributions, $\widehat{R}^\prime_\mathfrak{n}(t)$.}
\end{figure*}

\pagebreak
Objects,
\begin{equation} \label{eq:recon}
\widehat{\varepsilon}^\prime_{\mathfrak{m},\mathfrak{n}}(t) \equiv \mathfrak{f}^{(L)}_\mathfrak{m} \frac{\mel{0}{\Omega^\prime_\mathfrak{m}(t)\Omega_{\mathfrak{n}}^{\prime\dag}(0)}{0}}
{\mel{0}{\Omega^\prime_\mathfrak{n}(t)\Omega_{\mathfrak{n}}^{\prime\dag}(0)}{0}}\,,
\end{equation}
are analogous to Eq.~\ref{eq:x-talk_term} for the ratio $R^{\prime}_\mathfrak{n}(t)$, whenever $E_\mathfrak{n}> E_\mathfrak{m}$. It follows that we should be able to reconstruct as
\begin{equation*}
\widehat{R}^{\prime}_\mathfrak{n}(t) = \mathfrak{f}^{(L)}_\mathfrak{n} + \sum_{\mathfrak{m} < \mathfrak{n}} \widehat{\varepsilon}^\prime_{\mathfrak{m},\mathfrak{n}}(t) \, ,
\end{equation*}
and this will have the same time-dependence as $R^{\prime}_\mathfrak{n}(t)$ for times after the small-time excited state contributions have died out.

Comparisons between $R^\prime_{\mathfrak{n}}(t)$ and $\widehat{R}^\prime_{\mathfrak{n}}(t)$ for levels in the $[100]\, A_1$ irrep are shown in Fig.~\ref{fig:mel_KK_recon}. The values of $\mathfrak{f}^{(L)}_\mathfrak{m}$ for $\mathfrak{m} < \mathfrak{n}$ needed in Eq.~\ref{eq:recon} are obtained serially from timeslice fits to the lower-lying energy level correlation functions. 
Level 2 meets our criteria for having non-negligible late-time pollution since its $\mathfrak{f}^{(L)}$ value is significantly smaller than the corresponding values for levels 0 and 1, with the value for level 1 being over ten times larger in absolute value. We see that while the contribution of lower-lying states to levels 1 and 3 is negligible, for level 2, the contribution of the nominally suppressed level 1 is observably large. We also see that restricting timeslice fits to values of $t_\mathrm{max} \lesssim 1.5 t_0$ will reduce the impact of this pollution considerably.

\begin{widetext}

\section{Coupled-channel form factor parameterization}\label{sec:cc_fit_params}

To describe  timelike form factors in the coupled-channel region we use polynomial parameterizations for the smooth functions in Eq.~\ref{eq:fNovD},
\begin{equation}
\mathcal{F}_a(s)/m_\pi^2 = \sum_{n=0}^{N_a}\, h_{a,n}\cdot\qty(\frac{s-s_0}{s_0})^n\,,
\end{equation}
and we characterize each of them by the order of the polynomials, i.e.\ $\{N_{\pi\pi},N_{K\overline{K}}\}$, as listed in Tab.~\ref{tab:CC_funcforms}.

As an example, the result of the fit to $\mathcal{F}_{\mathfrak{n}}^{(L)}$ using the \emph{reference coupled-channel} scattering parameterization, and $N_{\pi\pi}=N_{K\overline{K}}=1$, has the following parameters:

\begin{center}
\footnotesize
    \begin{tabular}{rll}
    $h_{\pi\pi,0} = $                         & $-0.09\pm 0.60$   &
    \multirow{10}{*}{$\begin{bmatrix*}[r]1 & -0.5 & -0.5 & 0.6   \;\;\;\;\;      & -0.2 & 0.3  & -0.4     & 0.2  & {\bf -1} & -0.5 \\
                                           & 1    & -0.4 & 0.4   \;\;\; \;\;     & 0    & -0.6 & {\bf 1}  & -0.2 & 0.5      & {\bf 1}    \\
                                           &      & 1    & {\bf -1}  \;\;\;\;\;  & 0.1  & 0.3  & -0.5     & 0.1  & 0.5      & -0.5 \\
                                           &      &      & 1        \;\;\;\;\;   & -0.1 & -0.3 & 0.5      & -0.2 & -0.6     & 0.4  \\[2ex]
                                           &      &      &                       & 1    & -0.2 & 0        & -0.2 & 0.1      & 0    \\
                                           &      &      &                       &      & 1    & -0.6     & 0.6  & -0.2     & -0.6 \\
                                           &      &      &                       &      &      & 1        & -0.3 & 0.4      & {\bf 1}    \\
                                           &      &      &                       &      &      &          & 1    & -0.1     & -0.2 \\
                                           &      &      &                       &      &      &          &      & 1        & 0.5  \\
                                           &      &      &                       &      &      &          &      &          & 1  
                                        \end{bmatrix*}$}\\
    $h_{\pi\pi,1} = $                  & $0.001 \pm 0.400$   & \\
    $h_{\overline{K}K,0} = $        & $0.8 \pm 1.6$   & \\
    $h_{\overline{K}K,1} = $       & $-0.28 \pm 0.50$   & \\[2ex]
    $m = $                         & $0.1338\,(5) \cdot a_t^{-1}$ & \\
    $g_{\pi\pi}   = $                  & $0.441\,(9)$   & \\
    $g_{K\overline{K}} = $                  & $0.17\,(30)$   & \\
    $\gamma_{\pi\pi, \pi\pi} = $       & $(2.9 \pm 0.9) \cdot a_t^{2}$   & \\
    $\gamma_{\pi\pi, K\overline{K}} = $     & $-(2.4 \pm 5.0) \cdot a_t^{2}$   & \\
    $\gamma_{K\overline{K}, K\overline{K}} = $   & $-(2.2 \pm 4.0) \cdot a_t^{2}$   & \\[1.3ex]
    \multicolumn{3}{l}{\quad\quad\quad\quad $\chi^2/N_{\text{dof}}=\frac{126.9}{32-4}=4.53$\,,}
    \end{tabular}
\end{center}\vspace{-0.9cm}
\begin{equation}\label{eq:cc_ref_Fp}\end{equation}

where we present also the parameters of the scattering amplitude to illustrate the correlation between the functions $\mathcal{F}_a$ and $\mathcal{M}$~\footnote{The quoted $\chi^2$ describes only the variation of the smooth function parameters, $h$. The slight variations of the correlation matrix in Eq.~\ref{eq:cc_ref_Fp} with respect to what is reported in Eq.~\ref{eq:cc_ref} are due to the fact that only a subset of 348 configurations out of the 400 available to calculate the spectrum are used for the extraction of the form factors.}. Note that some of the smooth function parameters, $h$, are maximally correlated or anticorrelated with parameters in the scattering amplitude. 

Even though the smooth functions are individually consistent with zero, when weighted by the finite-volume correction factors or the scattering amplitude, the resulting values are not compatible with zero.
For this to occur, it is necessary, although not sufficient, that the functions $\mathcal{F}_a$ have a significant correlation with the scattering amplitude $\mathcal{M}$, which can be seen in Eq.~\ref{eq:cc_ref_Fp}.

\section{Spacelike form factor of the pion and renormalization constant}\label{sec:pi_3pt}

The pion form-factor in the \emph{spacelike} region was extracted from three-point correlation functions, $\langle 0 | \Omega_\pi(\Delta t) \, \mathcal{J}(t) \, \Omega^\dag_\pi(0) | 0 \rangle$, computed with a single value of $\Delta t=32\, a_t$. Details of the computational approach are presented in Ref.~\cite{Radhakrishnan:2022ubg} and Ref.~\cite{Shultz:2015pfa}. 
In order to cover a wide kinematic region, correlators were computed giving access to matrix elements, $\mel{\pi(\mathbf{p}_1)}{\mathcal{J}^i_{\rho,\text{lat}}}{\pi(\mathbf{p}_2)}$, for combinations of pion momenta up to $|\mathbf{p}_i|^2 \leq 6 \, \left(\tfrac{2\pi}{L}\right)^2$, and current momentum insertion up to $|\mathbf{p}_1-\mathbf{p}_2 |^2 \leq 4\, \left(\tfrac{2\pi}{L}\right)^2$.

In a previous analysis of some of these correlation functions in Ref.~\cite{Radhakrishnan:2022ubg}, the leading time dependence was removed by forming the combination,
\begin{equation} \label{eq:3pt_old}
\widetilde{C}_{\text{3pt}}(\Delta t, t) = 
\frac{\langle 0 | \Omega_\pi(\Delta t) \, \mathcal{J}(t) \, \Omega^\dag_\pi(0) | 0 \rangle}
{e^{-E_{\mathbf{p}_1}(\Delta t-t)}e^{-E_{\mathbf{p}_2}t}}\,,
\end{equation}
where $E_{\mathbf{p}}$ corresponds to the energy of a single-pion state of momentum $\mathbf{p}$, and where the normalization of the optimized operators follows the same convention as in the main text.

It can be the case that the timeslice-to-timeslice data correlation for this quantity is considerable, resulting in fits with reasonable values of $\chi^2$ which undershoot the data. One such case is presented in panel (a) of Fig.~\ref{fig:3pt}.

An alternative approach is to form a ratio using optimized two-point correlation functions,
%
\begin{equation} \label{eq:3pt_new}
R_{\text{3pt}}(\Delta t, t) = 
4E_{\mathbf{p}_1}E_{\mathbf{p}_2} \frac{\ev{\Omega_\pi(\Delta t) \, \mathcal{J}(t) \, \Omega^\dag_\pi(0) }{0}}
{\ev{\Omega_{\pi}(\Delta t \!-\! t)\,\Omega^\dag_{\pi}(0)}{0}\ev{\Omega_{\pi}(t)\,\Omega^\dag_{\pi}(0)}{0}}\,,
\end{equation}
%
which will have the same constant contribution, but differing excited-state contributions. This combination proves to have much smaller timeslice-to-timeslice data correlation, and fits follow more closely the data points. This is illustrated in panel (b) of Fig.~\ref{fig:3pt}. Fits to a constant, and a constant with an excited state exponential at source or sink or both are carried out for a range of time windows, and the results averaged using the AIC as in the two-point function case discussed previously. The columns on the right describe the time window of the fit $[t_{\text{min}},$ $t_{\text{max}}]$, and the number of exponentials at the source, $n_{\text{src}}$, and sink, $n_{\text{snk}}$.

\begin{figure}
\subfloat[\label{fig:pi_3pt_C}]{\includegraphics[width=.4\linewidth]{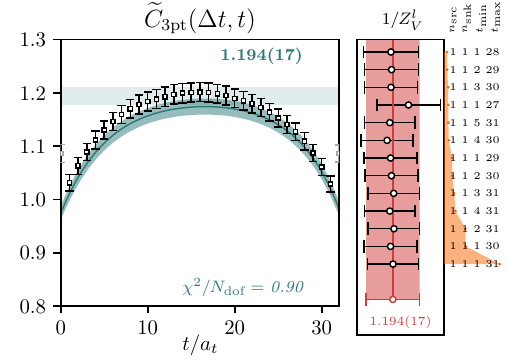}}
\hfil
\subfloat[\label{fig:pi_3pt_R}]{\includegraphics[width=.4\linewidth]{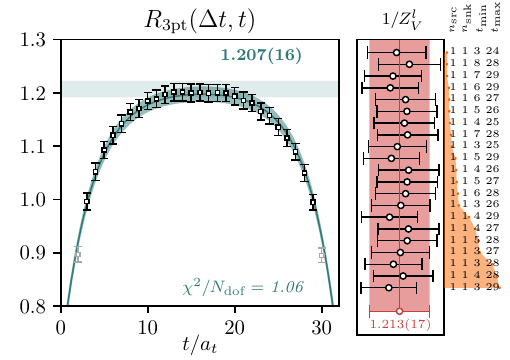}}
\caption{\label{fig:3pt}
Fits to three-point correlation functions with $\mathbf{p}_1 = \mathbf{p}_2 = \tfrac{2\pi}{L}[110]$ (averaged over rotations and directions of current insertion) and fixed $\Delta t = 32 \, a_t$ using either (a) Eq.~\ref{eq:3pt_old} or (b) Eq.~\ref{eq:3pt_new}. Fitted constant value in this case corresponds to $1/Z_V^\ell$, the vector current renormalization constant. Variation of fit window is shown in the right columns, along with the model-averaged result.
}
\end{figure}

\end{widetext}

\begin{figure}[h!]
\includegraphics[width=0.75\columnwidth]{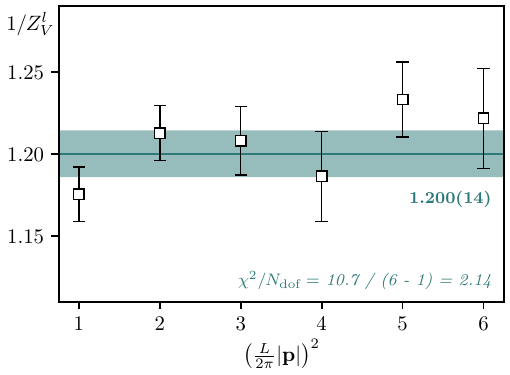}
\caption{Determination of light-quark vector current renormalization factor $Z_V^l$ by correlated fitting of extractions from six values of pion momentum.}
\label{fig:renorm}
\end{figure}

The difference with respect to the previous method using $\widetilde{C}_{\text{3pt}}(\Delta t, t)$ is modest, but is the origin of any differences in the current analysis with that in Ref.~\cite{Radhakrishnan:2022ubg}, such as for the light-quark vector current renormalization factor as shown in Fig.~\ref{fig:renorm}.

\end{document}